\documentclass[prx, 9pt,twocolumn,superscriptaddress]{revtex4-2}
\pdfoutput=1
\usepackage{bm}
\usepackage{bbm}
\usepackage{amsmath, amsfonts}
\usepackage{amssymb}
\usepackage{graphicx}
\usepackage{xcolor}
\usepackage[colorlinks=true,allcolors=blue]{hyperref}
\usepackage{cleveref}
\usepackage{mathtools}

\usepackage{newtxtext,newtxmath}

\usepackage[normalem]{ulem}
\begin{document}
\title{Fast $ZZ$-Free Entangling Gates for Superconducting Qubits Assisted by a Driven Resonator}
\author{Ziwen Huang}
\email{zhuang@fnal.gov}
\address{Superconducting Quantum Materials and Systems Center,
Fermi National Accelerator Laboratory (FNAL), Batavia, IL 60510, USA}
\author{Taeyoon Kim}
\address{Superconducting Quantum Materials and Systems Center,
Fermi National Accelerator Laboratory (FNAL), Batavia, IL 60510, USA}
\address{Department of Physics and Astronomy, Northwestern University, Evanston, IL 60208, USA}
\author{Tanay Roy}
\address{Superconducting Quantum Materials and Systems Center,
Fermi National Accelerator Laboratory (FNAL), Batavia, IL 60510, USA}
\author{Yao Lu}
\address{Superconducting Quantum Materials and Systems Center,
Fermi National Accelerator Laboratory (FNAL), Batavia, IL 60510, USA}
\author{Alexander Romanenko}
\address{Superconducting Quantum Materials and Systems Center,
Fermi National Accelerator Laboratory (FNAL), Batavia, IL 60510, USA}
\author{Shaojiang Zhu}
\address{Superconducting Quantum Materials and Systems Center,
Fermi National Accelerator Laboratory (FNAL), Batavia, IL 60510, USA}
\author{Anna Grassellino}
\email{annag@fnal.gov}
\address{Superconducting Quantum Materials and Systems Center,
Fermi National Accelerator Laboratory (FNAL), Batavia, IL 60510, USA}

\begin{abstract}
    Engineering high-fidelity two-qubit gates is an indispensable step toward practical quantum computing. For superconducting quantum platforms, one important setback is the stray interaction between qubits, which causes significant coherent errors. For transmon qubits, protocols for mitigating such errors usually involve fine-tuning the hardware parameters or introducing usually noisy flux-tunable couplers. In this work, we propose a simple scheme to cancel these stray interactions. The coupler used for such cancellation is a driven high-coherence resonator, where the amplitude and frequency of the drive serve as control knobs. Through the resonator-induced-phase ({\footnotesize RIP}) interaction, the static $ZZ$ coupling can be entirely neutralized. We numerically show that such a scheme can enable short and high-fidelity entangling gates, including cross-resonance {\footnotesize CNOT} gates within 40 ns and adiabatic {\footnotesize CZ} gates within 140 ns. Our architecture is not only $ZZ$ free but also contains no extra noisy components, such that it preserves the coherence times of fixed-frequency transmon qubits. With the state-of-the-art coherence times, the error of our cross-resonance {\footnotesize CNOT} gate can be reduced to below $10^{-4}$.
\end{abstract}

\maketitle
\section{Introduction}
Strong interactions are a typical requirement for faster entangling operations between multiple qubits
\cite{Blais_cqed_review,Oliver_scqubits_review}. However, increasing the coupling strength can come with larger stray interactions, which limit the fidelities of both single and two-qubit gates. For superconducting qubits, one solution is to introduce flux-tunable couplers, which can mitigate such cross-talk and has led to increased gate fidelities \cite{Oliver_tunable_coupling,Houck_ZZ_free,Houck_tunable_coupling,Oliver_two_qubit_gate,Yu_two_qubit_gates,Plourde_CR_tunable_coupler,Xu_weakly_tunable,Rigetti_tunable_coupler,Sun_tunable_coupler,IBM_tunable_coupling}. However, for the widely used transmon processors \cite{Koch_transmon_theory}, it has been observed that such tunable elements also lead to extra decoherence errors due to the high loss rate of the coupler \cite{rigetti_qpu,IBM_frequency_tunable_transmon,Paladino_TLF_review,Oliver_scqubits_review}.

Alternative strategies for $ZZ$ cancellation include using multi-path couplers \cite{IBM_direct_CR} and ac Stark shifts \cite{IBM_sizzle,IBM_Bgate,siddiqi_sizzle,Siddiqi_floquet_qubit}. By either fine-tuning the parameters of the bus coupler \cite{Zhao_bus_fine_tuning} or Stark driving the transmon qubits, one can suppress the $ZZ$ interaction without any flux tunability. In this paper, we propose a different non-tunable architecture that is even free of the parameter fine-tuning and direct transmon drives for the cancellation. In our scheme, a high-coherence microwave resonator is used to cancel unwanted $ZZ$ coupling via the resonator-induced-phase ({\footnotesize RIP}) interaction \cite{Blais_RIP_gate,IBM_RIP,IBM_RIP_leakage,IBM_RIP_optimization}. For such cancellation, we apply a constant off-resonant drive that displaces the resonator state ($\approx\!10$ photons in the steady state). Due to the dispersive shifts, the displacement of the resonator state depends on the states of the two transmon qubits, which in turn introduces a dynamical $ZZ$ interaction between the qubits. The strength of this coupling is highly tunable and can be employed to cancel the static $ZZ$ coupling. With superconducting microwave cavities with high coherence times \cite{Grassellino_Cavity,Rosenblum_cavity,Yale_2d_resonator}, the resonator photon loss only negligibly affects gate fidelities. 

We show by simulations that our architecture offers a family of fast entangling gates between qubits, including the cross-resonance ({\footnotesize CR}) \cite{Paraoanu_CR,Sheldon_CR,IBM_direct_CR,Devoret_Rigetti_CR,IBM_echo_CR} {\footnotesize CNOT} gates ($\lesssim\!40$ ns) and adiabatic {\footnotesize CZ} gates \cite{Oliver_two_qubit_gate,IBM_tunable_coupling} ($\lesssim\!140$ ns). Particularly, the {\footnotesize CR-CNOT} gates can be significantly accelerated since the cancellation of the static $ZZ$ coupling ($\approx$ 6 MHz in our simulation) allows stronger qubit coupling without introducing coherent errors caused by stray interactions \cite{IBM_sizzle}. Due to the shorter gate times, the infidelity of the {\footnotesize CR }gates can reach below $10^{-4}$ according to our simulation. Besides, the stronger coupling also allows a much larger qubit frequency separation compared to those on typical {\footnotesize CR }architectures \cite{Riverside_CR_errorbudget,IBM_CR_effective}, which can alleviate the issue of frequency crowding among the transmon qubits.

\begin{figure}[b]
    \centering
    \includegraphics[width = 6.5cm]{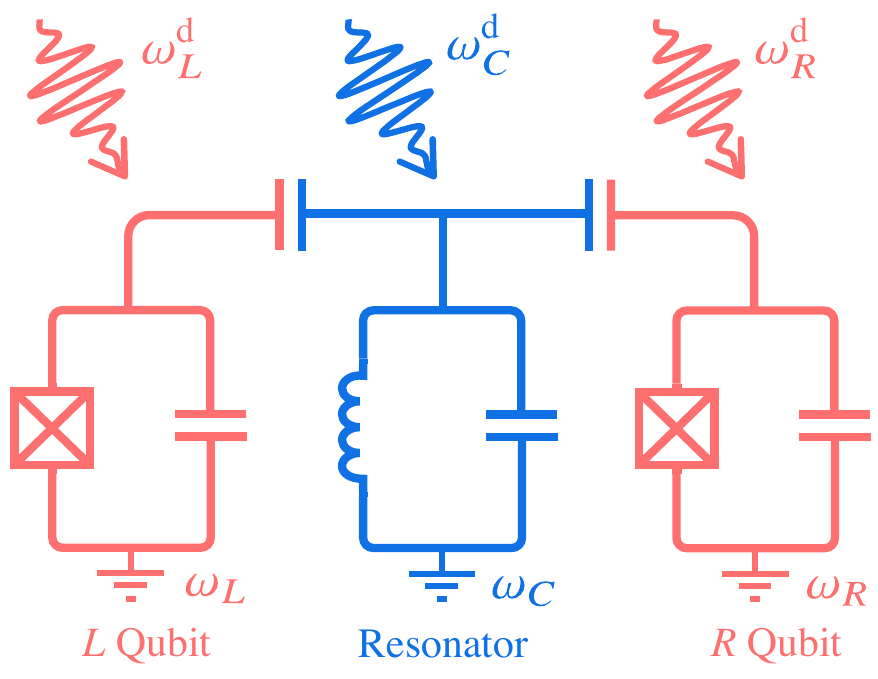}
    \caption{Schematic of the lumped-element model of the composite transmon-resonator-transmon system. In this model, both transmon qubits are capacitively coupled to a high-coherence resonator, which is modeled as a lumped-element resonator. The excitation energies of the left ($L$), right ($R$) qubit and the coupling resonator ($C$) are denoted by $\omega_{L,R,C}$, respectively. The frequencies of the drives on the three elements are denoted by $\omega^\mathrm{d}_{L,R,C}$.}
    \label{fig:cavity-qubits}
\end{figure}

The paper is structured as follows. In Sec. \ref{sec:model}, we introduce the lumped-element model of this architecture and demonstrate both analytically and numerically the $ZZ$ cancellation mechanism. In Sec. \ref{sec:CR}, we discuss the implementation of the fast cross-resonance gates and present simulations of their gate fidelities. In Sec. \ref{sec:CZ}, we show simulations of the adiabatic {\footnotesize CZ} gates via the adiabatic tuning of resonator drives. In Sec. \ref{sec:decoherence}, we briefly discuss other potential error channels and their magnitudes. In Sec. \ref{sec:scaleup}, we outline a plan to scale up this $ZZ$-free architecture. In Sec. \ref{sec:conclusion}, we append our conclusions.

\section{Model and $ZZ$ Cancellation}
\label{sec:model}
\subsection{Hamiltonian and dispersive transformation}

The circuit we study comprises two superconducting qubits both capacitively coupled to a central linear resonator.  The lumped-element model of this system is shown in Fig.~\ref{fig:cavity-qubits}. The Hamiltonian of this system is
\begin{align}
    \hat{H}(t) = \sum_{\upsilon=L,R}\hat{H}_{\upsilon} + \hat{H}_C+\hat{H}_{\mathrm{int}} + \hat{H}_\mathcal{D}(t),\label{eq:Hbarefull}
\end{align}
where $\hat{H}_{L(R)}$ denotes the Hamiltonian of the left (right) qubit, $\hat{H}_C$ is the resonator Hamiltonian, and $\hat{H}_{\mathrm{int}}$ is their interaction. Besides the static terms, the time-dependent Hamiltonian $\hat{H}_{\mathrm{d}}(t)$ describes the drives on the three components.

For this study, we consider the two qubits as fixed-frequency transmon qubits \cite{Koch_transmon_theory,Schoelkopf_3d_transmon,Houck_tantalum_qubit,Yu_tantalum_qubit,Fermilab_encapsulation,Roudsari_transmon_interfacial}, which are among the most coherent superconducting quantum elements and have been demonstrated to possess coherence times approaching one millisecond \cite{Houck_tantalum_qubit,Fermilab_encapsulation,Yu_tantalum_qubit,ibm_t1_fluc}. The Hamiltonian of the left (right) transmon qubit is given by $\hat{H}_{L(R)} = 4E_{C_{L(R)}}\hat{n}^2_{L(R)} - E_{J_{L(R)}}\cos\hat{\varphi}_{L(R)}$, where $\hat{n}_{L(R)}$ and $\hat{\varphi}_{L(R)}$ denote the charge and phase operators of the left (right)  qubit. The central lumped-element resonator is assumed to be physically implemented as a 2D or 3D high-coherence resonator. The Hamiltonian of this resonator is given by $\hat{H}_C = \bar{\omega}_C \hat{a}^\dagger \hat{a}$,
where $\hat{a}$ $(\hat{a}^\dagger)$ denotes the annihilation (creation) operator, and $\bar{\omega}_C$ is the bare resonator frequency. The interaction term is specified by $\hat{H}_{\mathrm{int}} = (\hat{a}+\hat{a}^\dagger)(g_L\hat{n}_L + g_R\hat{n}_R)$,  where $g_{L,R}$ denotes the qubit-resonator coupling strength. For these three components, the drive term $\hat{H}_\mathcal{D}(t)$ is specified as the charge drives that address each element. In the bare basis, it takes on the form
\begin{align}
    \hat{H}_{\mathcal{D}}(t) = \sum_{\upsilon=L,R}\mathcal{D}_{\upsilon}(t)\hat{n}_{\upsilon} + \mathcal{D}(t) (\hat{a}^\dagger + \hat{a}),
\end{align}
where $\mathcal{D}_{L,R}(t)$ and $\mathcal{D}(t)$ denote the time-dependent drive strengths.

Due to the coupling between the three components described by $\hat{H}_\mathrm{int}$, the true eigenstates of this composite system are hybridizations of the bare eigenstates of the Hamiltonians $\hat{H}_{L(R)}$  and $\hat{H}_C$. In terms of these true (dressed) eigenstates, the Hamiltonian in Eq.~\eqref{eq:Hbarefull} can be expressed as 
\begin{align}
    \hat{H}'(t) \approx &\!\!\!\sum_{\upsilon=L,R}\omega_{\upsilon}\hat{b}_{\upsilon}^\dagger\hat{b}_{\upsilon} + \frac{\eta_\upsilon}{2} \hat{b}_\upsilon^\dagger\hat{b}_\upsilon^\dagger\hat{b}_\upsilon\hat{b}_\upsilon + \chi_{\upsilon}\hat{b}_{\upsilon}^\dagger \hat{b}_{\upsilon} \hat{a}^\dagger\hat{a} \label{eq:dispersive}\\
    &+{\chi'_{LR}}\hat{b}_L^\dagger\hat{b}_L\hat{b}_R^\dagger\hat{b}_R+{\omega_C}\hat{a}^\dagger\hat{a}+\frac{\eta_C}{2}\hat{a}^\dagger\hat{a}^\dagger\hat{a}\hat{a}+    \hat{H}'_{\mathcal{D}}(t).\nonumber
\end{align}
Above, $\hat{b}_{L(R)}$ is the annihilation operator for the left (right) transmon qubit, $\omega_{L,R,C}$ denote the dressed frequencies of the left, right qubit, and the resonator, respectively. As common in experiments on fixed-frequency transmons \cite{IBM_RIP,IBM_parametric}, we consider the transmon frequencies to be around 4.5 GHz (detuning between qubit frequencies is specified later), and the frequency of the resonator is designed to be significantly detuned from those of the transmons $\omega_{L,R}$ to serve only as a passive coupler. (We consider this detuning $\approx 5$ GHz in this work.) The anharmonicity of the left (right) transmon qubit is denoted by $\eta_{L(R)}$. Because of the hybridization with the two non-linear transmon qubits, the resonator also acquires a small self-Kerr term in its Hamiltonian, whose strength is denoted by $\eta_C$. The full dispersive shift between the left (right) qubit and the resonator is denoted by $\chi_{L(R)}$. Due to their mutual coupling to the resonator and relatively closer frequencies, the two qubits also share a $ZZ$ interaction, whose magnitude is denoted by $\chi'_{LR}$. Finally, the drive Hamiltonian $\hat{H}'_{\mathcal{D}}(t)$ in Eq.~\eqref{eq:dispersive} is transformed from  $\hat{H}_{\mathcal{D}}(t)$ in Eq.~\eqref{eq:Hbarefull} after the dispersive treatment. We will specify it later in the following sections for different types of operations. 

In our simulations, we specify the dressed-basis Hamiltonian \eqref{eq:dispersive} by numerically diagonalizing the full Hamiltonian \eqref{eq:Hbarefull} via {\footnotesize SCQUBITS} \cite{Koch_scqubits}. The analytical expressions of the parameters in Eq.~\eqref{eq:dispersive} have been extensively explored in the literature, e.g., Refs.~\cite{IBM_CR_effective,IBM_RIP_leakage,Riverside_CR_errorbudget,IBM_CR_off_res_error}.

\subsection{Adiabatic elimination and $ZZ$ cancellation}

According to Eq.~\eqref{eq:dispersive}, the interaction between each qubit and the coupler resonator is described by the dispersive-shift terms such as $\chi_L\hat{b}^\dagger_L\hat{b}_L\hat{a}^\dagger\hat{a}$. If the passive resonator mode is kept in its ground state \cite{Sheldon_CR,IBM_direct_CR}, these interactions can be safely neglected. However, if we apply a drive on the resonator, such as \footnote{In the dressed basis, the charge operator of the resonator yields other small terms, but for a constant drive amplitude, those additional terms are far off-resonant.}
\begin{align}
    \hat{H}'_\mathcal{D,C}(t) = 2\mathcal{D}\cos\big(\omega^\mathrm{d}_Ct\big)(\hat{a}^\dagger+\hat{a}), \label{eq:cavity_drive}
\end{align}
this usually neglected interaction can induce entanglement between the two qubits \cite{Blais_RIP_gate,IBM_RIP}. Such interaction is the key element in this $ZZ$-free architecture. 

To see the $ZZ$ cancellation, we next derive an effective Hamiltonian with the resonator degree of freedom eliminated. After a frame transformation according to the unitary $\hat{U}_C(t) = \exp(-i\omega^\mathrm{d}_Ct\hat{a}^\dagger\hat{a})$, the  Hamiltonian in the rotating frame (with fast-rotating terms neglected) is given by
\begin{align}
    \tilde{H} = &\,\hat{U}^\dagger_C(t) \hat{H}(t) \hat{U}_C(t) - i{\hat{U}}^\dagger_C(t)\dot{\hat{U}}_C(t)\nonumber\\
    \approx &\,\hat{H}'_q-\Delta_d\hat{a}^\dagger\hat{a} + \frac{\eta_C}{2}\hat{a}^\dagger\hat{a}^\dagger\hat{a}\hat{a} + \mathcal{D}(\hat{a}^\dagger +\hat{a})\nonumber\\
    &+\chi_L\hat{b}_L^\dagger\hat{b}_L\hat{a}^\dagger\hat{a} +\chi_R\hat{b}_R^\dagger\hat{b}_R\hat{a}^\dagger\hat{a},\label{eq:rotate}
\end{align}
where we define $\Delta_d \equiv \omega^{\mathrm{d}}_C-\omega_C$, and $\hat{H}'_q$ is the Hamiltonian for two transmons [first line in Eq.~\eqref{eq:dispersive}]. Note that since the self-Kerr strength $\eta_C$ is much smaller than the other coefficients in the parameter regime we study, we neglect it for the following analytical derivation for simplicity but include it in our numerical simulation.

The drive term $\mathcal{D}(\hat{a}^\dagger +\hat{a})$ displaces the resonator state away from the zero-photon state. Therefore, we can no longer assume that the resonator only stays in its ground state. However, one can perform a transformation to adiabatically eliminate the drive term and further the resonator degree of freedom.  The unitary useful for such elimination is given by 
\begin{align}
    \hat{U}_{\mathrm{dis}} = \exp \Bigg[ \frac{\mathcal{D}(\hat{a}^\dagger-\hat{a})}{\Delta_d -\chi_L\hat{b}^\dagger_L\hat{b}_L-\chi_R\hat{b}^\dagger_R\hat{b}_R }\Bigg],
\end{align}
which further transforms Eq.~\eqref{eq:rotate} into
\begin{align}
    \tilde{H}_{\mathrm{dis}} = &\, \hat{U}^\dagger_{\mathrm{dis}} \tilde{H} \hat{U}_{\mathrm{dis}}\nonumber\\
    \approx &\,\hat{H}'_q+\big( -\Delta_d +\chi_L \hat{b}_L^\dagger\hat{b}_L +\chi_R \hat{b}_R^\dagger\hat{b}_R  \big)\hat{a}^\dagger\hat{a}\nonumber\\
    &+\frac{\mathcal{D}^2}{\Delta_d -\chi_L\hat{b}^\dagger_L\hat{b}_L-\chi_R\hat{b}^\dagger_R\hat{b}_R}.\label{eq:tildeHdisplace}
\end{align}
The second line in the equation above contains both the transmon Hamiltonians and the remaining coupling between the transmon qubit and the resonator in this displaced frame. The latter interaction can lead to shifts of the qubit frequencies if the resonator is not in the displaced vacuum state. However, if the resonator initially has zero photons and the ramp-up of the drive strength is sufficiently slow, during this ramp-up, the state of the resonator will remain in the vacuum defined in the displaced frame. Given that the resonator has a sufficiently small decoherence rate, we can safely assume that the resonator remains in this vacuum during our gate operations. (We discuss this effect in more detail in Sec. \ref{sec:decoherence}.)

\begin{figure}[t]
    \centering
    \includegraphics[width = 7.0cm]{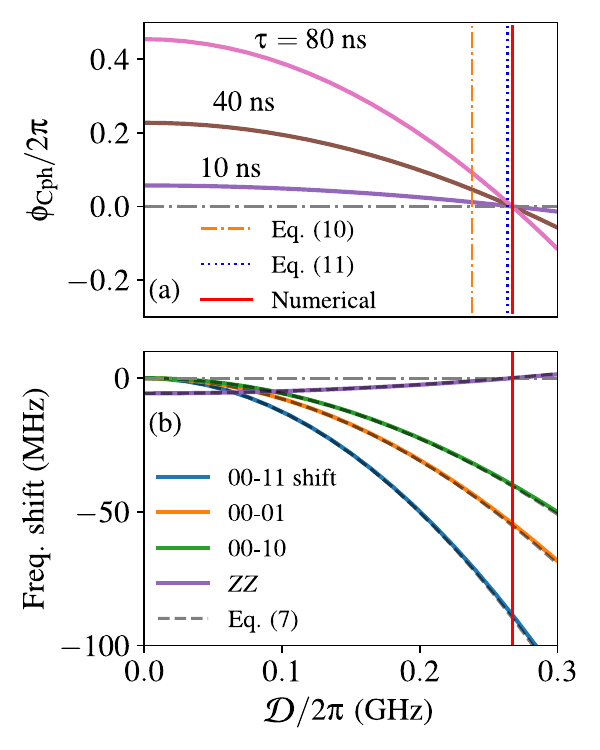}
    \caption{Numerical demonstration of $ZZ$ cancellation. Panel (a) plots the two-qubit controlled phase $\phi_{\mathrm{Cph}}$ as a function of the driving strength $\mathcal{D}$ for three different evolution times $\tau$. We confirm that $\phi_{\mathrm{Cph}}$ vanishes for all times for a certain $\mathcal{D}$, which is marked by the red solid line. Visibly, this value is close to the prediction by Eq.~\eqref{eq:ZZ_full}, but considerably differs from that by Eq.~\eqref{eq:ZZapproximation}. In (b), the energy shifts of different transition frequencies are plotted as functions of the driving strength. Dashed curves are predictions by Eq.~\eqref{eq:entangling_full}. The device parameters used for simulation are as follows. 
     The detuning of the two qubits is $\Delta_{LR}/2\pi  \approx -660$ MHz. The resonator drive frequency is detuned from the resonator frequency by $\Delta_d/2\pi = 100$ MHz. The anharmonicities of the three components are $\eta_L/2\pi\approx\eta_R/2\pi\approx-320$ MHz, and $\eta_C/2\pi\approx-90$ kHz. The three dispersive shifts are given by $\chi_{L}/2\pi \approx -6.0$ MHz, $\chi_R/2\pi \approx -8.4$ MHz, and $\chi'_{LR}/2\pi = -5.7$ MHz. The effective $ZZ$ coupling between the qubits vanishes at $\mathcal{D}_0/2\pi\approx$ 0.27 GHz.}
    \label{fig:czcancel}
\end{figure}

The third line in Eq.~\eqref{eq:tildeHdisplace} describes the {\footnotesize RIP }interaction between the two qubits introduced by the driven resonator. Approximately, in the limit of $|\chi_L|,|\chi_R|\ll |\Delta_d|$, the strength of the dynamical $ZZ$ coupling can be evaluated by expanding the last line of Eq.~\eqref{eq:tildeHdisplace} by
\begin{align}
    &\, \frac{\mathcal{D}^2}{\Delta_d - \chi_{L} \hat{b}_{L}^\dagger \hat{b}_{L} - \chi_{R} \hat{b}_{R}^\dagger \hat{b}_{R}}\label{eq:entangling_full}\\
    \approx&\,
    \frac{\mathcal{D}^2}{\Delta_d} + \frac{\mathcal{D}^2}{\Delta_d^2}\Big( \chi_{L} \hat{b}_{L}^\dagger \hat{b}_{L} + \chi_{R} \hat{b}_{R}^\dagger \hat{b}_{R}\Big)\nonumber\\
    &-\frac{\mathcal{D}^2}{\Delta_d^3}\Big( \chi_{L} \hat{b}_{L}^\dagger \hat{b}_{L} + \chi_{R} \hat{b}_{R}^\dagger \hat{b}_{R}\Big)^2\label{eq:ZZapproximation}.
\end{align}
Eq.~\eqref{eq:ZZapproximation} results in a two-qubit coupling term $-2\mathcal{D}^2\chi_L\chi_R\hat{b}^\dagger_{L}\hat{b}_{L}\hat{b}^\dagger_{R}\hat{b}_{R}/\Delta_d^3$ \cite{Blais_RIP_gate}, which provides a sign and amplitude-tunable $ZZ$ interaction. The condition 
\begin{align}
2\chi_L\chi_R\mathcal{D}^2/\Delta_d^3 = \chi'_{LR}\label{second-order-condition}
\end{align}
then approximately gives the $ZZ$-free operating point. However, this approximation only holds in the limit of  $|\chi_{L,R}|\ll |\Delta_d|$. In Appendix \ref{sizzle_approach}, we show another approach to derive the expression of the magnitude of this dynamical coupling.

Beyond this limit, we find that using Eq.~\eqref{eq:entangling_full} to calculate the entangling strength is more accurate. Specifically, the $ZZ$ cancellation requires 
\begin{align}
    \Delta E_{\mathrm{zp}}(1,1)\! + \Delta E_{\mathrm{zp}}(0,0)\! -  \Delta E_{\mathrm{zp}}(1,0)\! - \Delta E_{\mathrm{zp}}(0,1)\! + \chi'_{LR} = 0,\label{eq:ZZ_full}
\end{align}
where we define 
\begin{align}
    \Delta E_{\mathrm{zp}}(j_L,j_R)\equiv \frac{\mathcal{D}^2}{\Delta_d - j_L\chi_L-j_R\chi_R}.
\end{align}

To numerically confirm such a prediction, we perform a simulation \cite{qutip} of the controlled phase in a transmon-resonator-transmon system described by Hamiltonian \eqref{eq:dispersive}, where the previously neglected resonator self-Kerr term is included. We present our results in Fig.~\ref{fig:czcancel} (a). For this simulation, we choose realistic parameters motivated by previous experiments \cite{Grassellino_Cavity,Rosenblum_cavity,IBM_RIP}, which are listed in the caption. The controlled phase is obtained by evaluating $\phi_{\mathrm{Cph}} = \phi_{10}+\phi_{01} -\phi_{00}-\phi_{11}$ for a given evolution time $\tau$ and resonator driving strength $\mathcal{D}$, where $\phi_{j_L,j_R}$ denotes the system's accumulated phase if the number of photons in the left (right) transmon qubit is maintained at $j_L(j_R)$. By varying $\mathcal{D}$ and $\tau$, we find that $\phi_{\mathrm{Cph}}$ vanishes at all times for a certain $\mathcal{D}$, implying the desired $ZZ$ cancellation. Such a value considerably differs from the leading-order approximation \eqref{eq:ZZapproximation}, but can be closely predicted by solving Eq.~\eqref{eq:ZZ_full} for $\mathcal{D}$. The remaining deviation between the prediction by Eq.~\eqref{eq:ZZ_full} and the numerical result is related to the neglection of the weak resonator self-Kerr described by $\eta_C\hat{a}^\dagger\hat{a}^\dagger\hat{a}\hat{a}/2$. 

Besides canceling the $ZZ$ interaction, turning on a resonator drive also shifts the 0-1 excitation frequencies of the two qubits. In Fig.~\ref{fig:czcancel} (b), we also plot the shift of the 0-1 transition frequencies. These shifts are also well approximated by Eq.~\eqref{eq:entangling_full}. If the strength of the drive on the resonator is sufficiently stable, these frequency shifts can be easily calibrated experimentally.

\section{Fast Cross-Resonance Gate}
\label{sec:CR}

For such a fixed-frequency architecture, one of the most popular and convenient gates to apply is the cross-resonance gate \cite{Paraoanu_CR,Devoret_CR_gate,Riverside_CR_errorbudget,IBM_CR}. Generally, this type of entangling gate takes longer times than parametric gates on frequency tunable architectures \cite{Oliver_two_qubit_gate,Yu_two_qubit_gates,google_quantum_superamacy,Schuster_qutrit,rigetti_qpu,IBM_tunable_coupling}. Impressively, we find our $ZZ$-free scheme can significantly reduce the required gate time, which can be comparable to those obtained for the parametric gates.

\subsection{Traditional {\footnotesize CR }gates and their errors}
To see how the $ZZ$ cancellation can help with improving the fidelities of the {\footnotesize CR }gates, we first revisit the traditional {\footnotesize CR }gates, and understand the current limitations on gate fidelities. It has been pointed out that three factors especially contribute to gate errors \cite{Riverside_CR_errorbudget, IBM_CR_off_res_error,IBM_direct_CR,IBM_sizzle,IBM_CR_off_res_error}, which are summarized and discussed below.

\textit{Decoherence.--} We assume that the strength of the coupling between the two transmon qubits is $J$, the detuning between the frequencies of the two qubits is $\Delta_{LR}\equiv \omega_L-\omega_R$, and their anharmonicities satisfy $\eta_L\approx\eta_R\approx\eta$. For the resonator mediated coupling assumed in our architecture, the coupling strength $J$ is approximately \cite{IBM_RIP_leakage,IBM_CR_effective}
\begin{align}
    J\approx\, \tilde{g}_L \tilde{g}_R\Bigg[& \frac{\bar{\omega}_L+\bar{\omega}_R-2\bar{\omega}_C}{2(\bar{\omega}_L-\bar{\omega}_C)(\bar{\omega}_R-\bar{\omega}_C)}\nonumber\\
     -&\frac{\bar{\omega}_L+\bar{\omega}_R+2\bar{\omega}_C}{2(\bar{\omega}_L+\bar{\omega}_C)(\bar{\omega}_R+\bar{\omega}_C)}  \Bigg],
\end{align}
where $\bar{\omega}_{L,R,C}$ are the bare frequencies of the two qubits and the resonator, and $\tilde{g}_{L(R)}$ are the normalized coupling strength defined by $\tilde{g}_{L(R)}\approx [E_{J_{L(R)}}/32E_{C_{L(R)}}]^{\frac{1}{4}}g_{L(R)}$. To activate the cross-resonance gate, we consider a drive on the control qubit (chosen as $L$) whose frequency is close to that of the target qubit ($R$). The strength of this drive is denoted by $\epsilon_{\mathrm{CR}}$.

Generally speaking, a longer gate time implies more decoherence loss \cite{Goran_universal_reduction,Huang_CPTP}. To reduce the {\footnotesize CR }gate duration, a stronger effective {\footnotesize CR }driving strength is needed. The effective $CX$ rotation rate induced by the {\footnotesize CR }drive is approximately $\epsilon_{CX}\approx A_{CX}\epsilon_{\mathrm{CR}}$, where $\epsilon_{CR}$ is the amplitude of the drive on the control qubit, and the coefficient is approximately \cite{Riverside_CR_errorbudget,Blais_cqed_review}
\begin{align}
    A_{CX}\approx \Bigg(\frac{E_{J_{L}}}{32E_{C_{L}}}\Bigg)^{\frac{1}{4}}\frac{2J\eta}{\Delta_{LR} (\eta+\Delta_{LR})}.\label{eq:ecr}
\end{align}
Roughly, if the transmon loss rate is characterized by $\gamma$, the decoherence error scales as
\begin{align}
\mathrm{Err}_{\mathrm{noise}} \approx&\, \frac{4\pi\gamma}{5|\epsilon_{CX}|} \nonumber\\
\propto & \begin{cases}
        {|\Delta_{LR}}| \cdot|J|^{-1}\cdot|\epsilon_{\mathrm{CR}}|^{-1}\cdot \gamma, &|\Delta_{LR}|\ll |\eta|;\\
        |\Delta_{LR}| ^2\cdot|J|^{-1}\cdot|\epsilon_{\mathrm{CR}}|^{-1}\cdot |\eta|^{-1} \cdot \gamma,  & |\Delta_{LR}|\gg |\eta|.
\end{cases}
\end{align}

\textit{Leakage.--} The second major error source is the leakage due to off-resonant transitions. Ideally, the {\footnotesize CR }drive only induces rotations in the target qubit; however, to ensure a sufficiently fast gate, the control qubit is strongly driven, as shown in Eq.~\eqref{eq:ecr}. Such drive causes off-resonant errors \cite{Riverside_CR_errorbudget,IBM_CR_off_res_error}. An accurate estimation of such error requires the knowledge of specific parameter regimes of the device and the pulse shapes \cite{Riverside_CR_errorbudget,IBM_CR_off_res_error}, but generally speaking, this off-resonant error $\mathrm{Err}_{\mathrm{leak}}$ increases with larger driving strength $\epsilon_{\mathrm{CR}}$ but decreases with larger detuning $\Delta_{LR}$.

\textit{Imperfect rotation.--} Due to the unwanted interactions, especially the $ZZ$ coupling, the conditional Rabi rotation usually cannot be made fully resonant. To quantify this type of error, we first evaluate the $ZZ$ rate, which is approximated by \cite{IBM_direct_CR,Riverside_CR_errorbudget}
\begin{align}
    \chi'_{LR} \approx \frac{4J^2\eta}{(\Delta_{LR}+\eta)(\Delta_{LR}-\eta)}.
\end{align}
Such coupling introduces a coherent error roughly proportional to \cite{IBM_direct_CR,Riverside_CR_errorbudget}
\begin{align}
    \mathrm{Err}_{ZZ}\propto &\,\Big|\frac{\chi^{\prime}_{LR}}{\epsilon_{CX}}\Big|^2\nonumber\\
    \propto& \begin{cases}
        |\Delta_{LR}|^2\cdot|J|^2\cdot|\epsilon_{\mathrm{CR}}|^{-2}\cdot |\eta|^{-2}, &|\Delta_{LR}|\ll |\eta|;\\
        |J|^2\cdot|\epsilon_{\mathrm{CR}}|^{-2}, &|\Delta_{LR}|\gg |\eta|.
    \end{cases}
\end{align}

\begin{figure}
    \centering
    \includegraphics[width = 6.5cm]{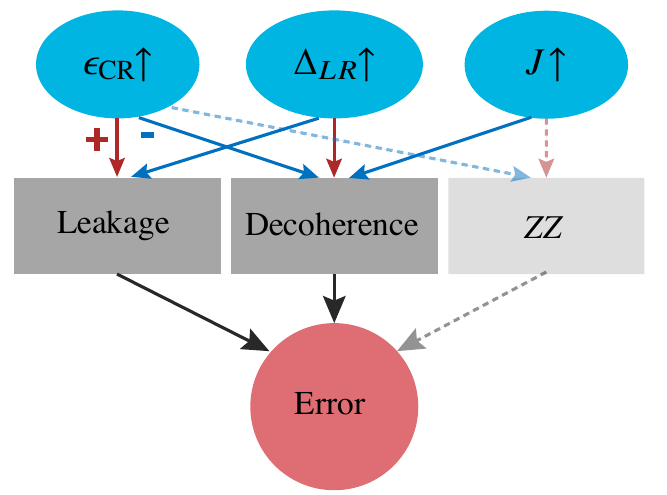}
    \caption{A diagram illustrating how traditional cross-resonance gates are affected by the three error channels, namely the decoherence loss, leakage, and imperfect rotations due to $ZZ$ interactions. The contributions of these channels are determined by three parameters, i.e., the two-qubit detuning $\Delta_{LR}$, the driving strength $\epsilon_\mathrm{CR}$, and the two-qubit effective coupling strength $J$. The dependence is roughly characterized by positive (red arrows) and negative (blue arrows) correlations, as sketched in the diagram above. Using the architecture presented in this work, the $ZZ$ error can be removed. Such mitigation strategy is indicated by the dashed lines.}\label{fig:CRerrordiagram}
\end{figure}

By inspecting the scaling laws of the magnitude of the three types of errors, one can find that there is no clear parameter regime where all three can be suppressed simultaneously \cite{IBM_sizzle}. Specifically, reducing $\mathrm{Err}_{\mathrm{noise}}$ requires a smaller $\Delta_{LR}$, reducing  $\mathrm{Err}_{\mathrm{leak}}$ requires a smaller $\epsilon_\mathrm{CR}$, and reducing $\mathrm{Err}_{ZZ}$ requires a smaller $J$. However, if all of them are small, neither type of error can be efficiently suppressed. We summarize this situation in a diagram in Fig.~\ref{fig:CRerrordiagram}.

To overcome this apparent ``trilemma", one promising route is to cancel or suppress the $ZZ$ coupling strength. This has been achieved by introducing extra coupling elements \cite{IBM_direct_CR,Plourde_CR_tunable_coupler} or additional drives \cite{IBM_sizzle}. These cancellation approaches open up new parameter space for error reduction. For example, if we maintain a reasonable ratio $\epsilon_{\mathrm{CR}}/\Delta_{LR}$ to suppress leakage, we can increase the coupling strength $J$ between the two qubits to further reduce the gate infidelity \cite{IBM_sizzle}.

In our $ZZ$-free regime,  the third error source is eliminated by the {\footnotesize RIP }interaction, which does not introduce additional loss channels, or require the fine-tuning of hardware parameters. The drive is applied to the resonator, which can avoid directly affecting the coherence times of the transmon qubits \cite{Yale_bilinear_coupling,Yao_beam_splitting}. The highly tunable dynamical $ZZ$ interaction provides a tool to cancel a relatively stronger static $ZZ$ coupling. In the next subsection, we show our simulation where we implement {\footnotesize CR }gates with the static $ZZ$ coupling at
$\chi'_{LR}/2\pi\approx-5.7$ MHz dynamically cancelled by the resonator drive.

\subsection{$ZZ$-free {\footnotesize CR }gate}
Motivated by the consideration above, we choose a stronger two-qubit coupling strength (effectively $|J/2\pi|\approx 42$ MHz)  compared to the values used in the literature \cite{IBM_direct_CR,IBM_CR,IBM_sizzle}. Such strong coupling allows us to explore beyond the straddling regime, where the problem of frequency crowding between qubits can be alleviated. For example, we choose the detuning as $\Delta_{LR}/2\pi= -660$ MHz. These choices result in a static $ZZ$ coupling strength $\chi'_{LR}/2\pi = -5.7$ MHz. Other parameters are listed in the caption of Fig.~\ref{fig:czcancel}. To neutralize the stray coupling, we operate the resonator at the $ZZ$ free point $\mathcal{D}=\mathcal{D}_0$ as shown in Fig.~\ref{fig:czcancel} (a). 

\begin{figure}[h!t]
    \centering
    \includegraphics[width = 8.6cm]{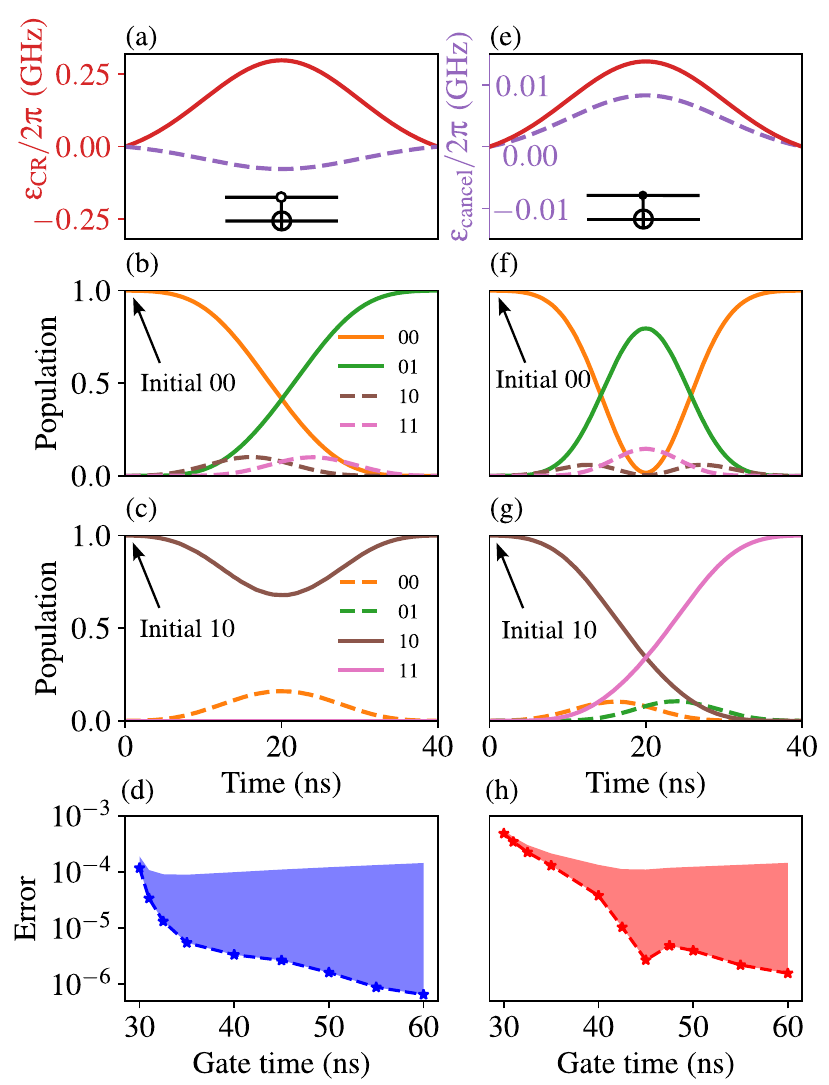}
    \caption{Numerical simulation of the cross-resonance gates with $ZZ$ cancellation. Two types of gates are chosen for such simulation: the left panels (a)-(d) show the results for the 0-Controlled {\footnotesize NOT} gate and the right panels (e)-(h) for the 1-Controlled {\footnotesize NOT}, respectively. Among these panels, (a) and (e) show the envelopes of the {\footnotesize CR }drive on the control qubit (red solid) and the cancellation drive (purple dashed) on the target qubit, and their insets contain circuit diagrams for the two types of gates. The envelopes are chosen as truncated Gaussian functions ($2\sigma$ on each side). Panels (b), (c), (f), and (g) show evolutions of the four computational states during 40-ns gates. (d) and (h) present the minimized coherent gate error for the two types of entangling gates versus given gate times. For this optimization, we fix $\omega^\mathrm{d}_R=\omega^\mathrm{d}_L$, and vary the driving frequency $\omega^\mathrm{d}_L$, the maximal amplitude of $\epsilon_{\,\mathrm{CR}}(t)$, and the maximal amplitude of $\epsilon_{\,\mathrm{cancel}}(t)$ to search for the minimal average gate error. As shown in stars in (d) and (h), the coherent errors (shown in blue and red stars) are below $10^{-4}$ for most gate durations simulated. Consequently, the decoherence contribution dominates the total gate error. Even for an optimistic estimation of the transmon coherence times ($T_1=T_2=500\,\mu$s), the decoherence errors still predominate the coherent ones for gate durations longer than 40 ns, which is indicated by the shaded areas. (We assume that both the depolarization and dephasing noise are Markovian.) For all the simulations, the strength of the drive on the resonator is fixed at the $ZZ$-free point, i.e., $\mathcal{D} = \mathcal{D}_0$.}
    \label{fig:CR}
\end{figure}

To verify the viability of the {\footnotesize CR }gates, we use the model described in Eq.~\eqref{eq:dispersive} to numerically simulate the evolution of the qubit states under {\footnotesize CR }gate drives. For this simulation, we choose the $L$ qubit as the control and $R$ as the target, and  drive the $L$ qubit via the charge operator $\hat{n}_L$ according to 
\begin{align}
    \hat{H}_{\mathcal{D},L}(t) = 2\epsilon_{\mathrm{CR}}(t)\cos\big(\omega^\mathrm{d}_L t\big)\hat{n}_L.
    \label{eq:transmondrive}
\end{align}
Importantly, the bare charge operator is transformed into a sum of contributions in the dressed basis, which is expressed as
\begin{align}
    \hat{n}_L \rightarrow &\,A_{L}\hat{b}_L + A_{R}\hat{b}_R + A_{CX} \hat{b}^\dagger_L\hat{b}_L\hat{b}_R \nonumber\\
    &+A'_{R}\hat{b}^\dagger_R\hat{b}_R\hat{b}_R +  A'_{CX} \hat{b}^\dagger_L\hat{b}_L\hat{b}^\dagger_R\hat{b}_R\hat{b}_R+\cdots\nonumber\\
    &+\text{H.c.}\label{eq:nexpand}
\end{align}
Again, we obtain the coefficients in this expansion numerically using {\footnotesize SCQUBITS}. Among the terms in this expansion, the following are important for understanding and designing the {\footnotesize CR} gate. Specifically, $A_L\hat{b}_L$ describes the drive on the control qubit, and $A_{R}\hat{b}_R$ induces single-qubit rotations on the target qubit. Their coefficients can be approximated as $A_L\approx \big(E_{J_L}/32E_{C_L}\big)^{\frac{1}{4}}$, and $A_R\approx \big(E_{J_L}/32E_{C_L}\big)^{\frac{1}{4}} (-J/\Delta_{LR})$. The term $A_{CX} \hat{b}^\dagger_L\hat{b}_L\hat{b}_R$ induces conditioned rotation depending on the state of the control qubit. [See approximation of $A_{CX}$ in Eq.~\eqref{eq:ecr}.] This term can be put on resonance if the drive frequency $\omega^\mathrm{d}_L$ is chosen to be close to the frequency of the target qubit $\omega_R$. The other two terms shown in Eq.~\eqref{eq:nexpand} are related to the possible leakage in this gate, which we aim to minimize. They are also included in our following simulation.

Besides the drive on the resonator and on the control qubit, we also apply a cancellation pulse to the target qubit \cite{IBM_direct_CR} to directly correct the rotations in the target qubit induced by the term $A_R\hat{b}_R$. This drive is denoted by \begin{align}
    \hat{H}_{\mathcal{D},R}(t) = 2\epsilon_{\mathrm{cancel}}(t)\cos\big(\omega^\mathrm{d}_Rt\big)\hat{n}_R.
\end{align}
By tuning and the strength and phase of $\epsilon_{\mathrm{cancel}}(t)$, 
this cancellation tone can activate either a 0-{\footnotesize CNOT} gate (target qubit flipped when the control is in the ground state) or a 1-{\footnotesize CNOT} gate [see circuit diagrams in the insets of Fig.~\ref{fig:CR} (a) and (e)]. They are equivalent to each other up to single-qubit rotations. The envelopes of these two pulses  $\epsilon_{\mathrm{CR}}(t)$ and $\epsilon_{\mathrm{cancel}}(t)$ are shown in Fig.~\ref{fig:CR} (a) and (e). Besides these two, we also add DRAG pulses \cite{Wilhelm_DRAG,Wilhelm_DRAG_analy} in our simulations to mitigate leakage to the second excited state of the target qubit.

By fine-tuning the pulse parameters, we numerically realize the two types of entangling gates with gate times ranging from $30$ to $60$ ns. For a clearer demonstration of this gate, we 
plot the evolution of the populations of the four computational states during a 40-ns gate in Fig.~\ref{fig:CR} (b) and (c) for the 0-controlled {\footnotesize CNOT} gate, and in (f) and (g) for the 1-controlled {\footnotesize CNOT} gate. Clearly, the target qubit ends up in different states for different initial states of the control qubit. In Fig.~\ref{fig:CR} (d) and (h), we show the optimized gate fidelites for different gate times. Impressively, the coherent error is negligible ($\approx 10^{-5}$) for gate times as short as $40$ ns.

For such low coherence errors, the two-qubit gates are still limited by the transmon decoherence. For the state-of-the-art coherence times of the transmon qubits ($T_1=T_2=500 \,\mu s$) \cite{Houck_tantalum_qubit,Yu_tantalum_qubit,Fermilab_encapsulation}, the gate error of this gate is below $ 10^{-4}$ for a 40-ns duration, which is shown in Fig.~\ref{fig:CR} (d) and (h). Even for more moderate coherence times such as $T_1=T_2=100 \,\mu s$, this error can still reach as low as $5\times 10^{-4}$. Such error can be further reduced once the coherence times of single transmon qubits are further improved, and the absence of flux-tunable elements spares the gates from suffering from other prominent decoherence channels.

\section{Adiabatic CZ gates}
\label{sec:CZ}

Besides the {\footnotesize CR }gates, this architecture supports other entangling operations, which can complement the gates introduced above.  For example, we can engineer {\footnotesize CZ} gates via the {\footnotesize RIP }interaction by adiabatic tuning the resonator driving strength. This strategy is analogous to that used in gates enabled by tunable couplers \cite{IBM_tunable_coupling}. 

\begin{figure}[t]
    \centering
    \includegraphics[width = 7.3cm]{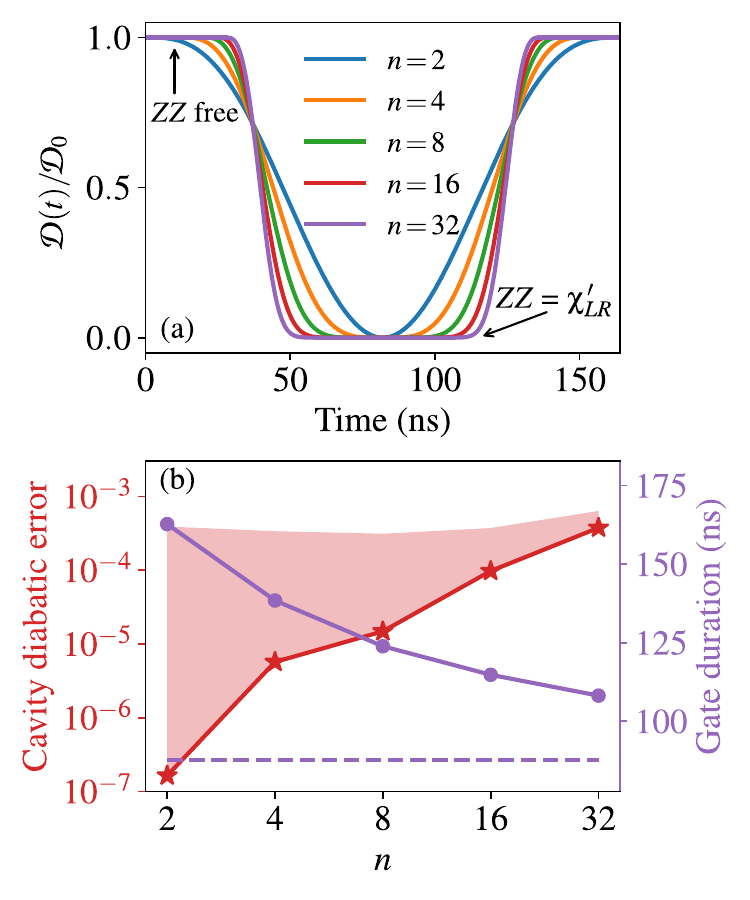}
    \caption{Numerical simulations of the adiabatic {\footnotesize CZ} gates. (a) shows the envelopes of the resonator drives that enable such phase entangling. The different curves correspond to different parameters $n$ in Eq.~\eqref{eq:f(t)}. (b) shows the optimized gate durations and corresponding resonator diabatic errors as a function of $n$. The dashed curve indicates the lower bound of gate duration set by $T_\mathrm{min} = \pi/{\chi'_{LR}}$. The shaded region indicates the decoherence contribution for transmon qubits with $T_1=T_2=500 \,\mu$s.}
    \label{fig:CZ}
\end{figure}

Specifically, we adiabatically move $\mathcal{D}$ in Eq.~\eqref{eq:cavity_drive} away from the cancellation point $\mathcal{D}_0$ over a certain duration $T_g$ and return it back. The  $ZZ$ coupling is recovered during this process, which can be used to engineer {\footnotesize CZ} gates. To maximally accelerate the entangling gate, we adiabatically tune $\mathcal{D}$ to 0 in our simulation [see Fig.~\ref{fig:CZ} (a)]. To ensure high-fidelity state transfers, we choose an analytically designed pulse envelope \cite{Muga_adiabaticity_shortcut}
\begin{align}
    \mathcal{D}(t) = 2^n\mathcal{D}_0\Bigg[ 10\left(\frac{t}{T_g}\right)^3 - 15\left(\frac{t}{T_g}\right)^4 + 6 \left(\frac{t}{T_g}\right)^5 -\frac{1}{2}\Bigg]^n,\label{eq:f(t)}
\end{align}
whose first and second-order derivatives vanish at $t=0$ and $t=T$ for $n=2m$, $m\in\mathbb{Z}^+$. Such a feature is useful for preserving adiabaticity in the state evolution of a harmonic oscillator. 

By increasing $n$, one can shorten the ramp time of the pulse as shown in Fig.~\ref{fig:CZ} (a), which leads to faster {\footnotesize CZ} gates at the cost of more diabatic leakage. To verify this, we increase $n$ from 2 to 32 and find optimal gate durations for the target {\footnotesize CZ} gates. As shown in Fig.~\ref{fig:CZ} (b), the gate duration decreases from 160 ns to 110 ns, while the error due to diabatic resonator evolution increases from a negligible value to $4\times 10^{-4}$, approaching the decoherence limit.  In addition, shorter pulse duration eventually results in a broader bandwidth of the pulse, which increases the risk of photon exchanges between the qubits and the resonator and leads to more errors \cite{IBM_RIP_leakage,Martinis_leakage}. In principle, this problem can be alleviated by carefully choosing the hardware and driving parameters. We leave the analysis and mitigation of these error channels in the current scheme for future investigation.

Although the durations of such {\footnotesize CZ} gates are relatively longer than those of the {\footnotesize CR }gates studied previously, they require simpler control protocols, which are especially convenient for experiments where strong transmon drives are unavailable.

\section{Other error channels}
\label{sec:decoherence}
\subsection{Resonator Decoherence}

In the previous sections, we have neglected the decoherence of the resonator by assuming that they have exceedingly longer coherence times than the transmon qubits \cite{Grassellino_Cavity,Rosenblum_cavity,Yale_2d_resonator}. However, such a condition may not always apply in the experiments. In this subsection, we quantify the error caused by the decoherence of the resonator.

For a fixed resonator drive, the resonator-induced dephasing for the $L$ transmon qubit is given by \cite{Schoelkopf_driven_cavity_dephasing}

\begin{align}
    \Gamma_{m,L} \approx \frac{2\bar{n}\kappa_C \chi^2_{L}}{\kappa_C^2 + \chi^2_L + 4\Delta_d^2},\label{eq:decoherence}
\end{align}
where $\kappa_C$ denotes the resonator loss rate and $\bar{n}$ denotes the average photon number kept in the resonator. For the control protocols we consider in this paper, the detuning $\Delta_d$ is much larger than $\chi_L$ and $\kappa_C$. Under the limit $\Delta_d\gg\chi_L,\kappa_C$,  Eq.~\eqref{eq:decoherence} can be further approximated as $\Gamma_{m,L}\approx (\bar{n}\chi^2_L/2\Delta_d^2)\kappa_C$. For a simple estimation, we take the parameters used in the previous simulations, i.e., $\Delta_d/2\pi = 100$ MHz, $\chi_L/2\pi = 6$ MHz, and $\bar{n}\approx 10$, and find the coefficient $\bar{n}\chi_L^2/2\Delta_d^2\approx 0.02$. Therefore, for the resonator lifetime $1/\kappa_C = 100\,\mu$s, the coherence limit on the transmon qubit is approximately 5 ms, which is still much higher than its typical dephasing time.

For non-steady drives applied to the resonator such as those proposed in Sec.~\ref{sec:CZ}, the simple estimation \eqref{eq:decoherence} should be modified by a more detailed formula provided in Ref.~\cite{Blais_RIP_gate}. However, we deem that such modification does not qualitatively change the estimation given above. Besides the inherent loss of the resonator, the stability of the microwave drives on the resonator can also be an error source that reduces the transmon coherence times. Since the magnitude of such instability has not been well established, we choose to leave such a contribution for future research.

\subsection{Four-wave mixing}

For the traditional {\footnotesize CR }gates \cite{IBM_CR_effective}, the charge drives on either transmon qubit usually only affect the qubits, since the frequencies of these drives are far detuned from that of the resonator excitation. In our scheme, however, if a displacement drive Eq.~\eqref{eq:cavity_drive} is applied to the resonator, the transmon drive Eq.~\eqref{eq:transmondrive} can introduce an intriguing four-wave-mixing interaction between the qubit and resonator.

We can make this interaction more evident by inspecting the transformation of $\hat{b}_{L(R)}$ according to the unitary $\hat{U}_{\mathrm{dis}}$. Toward this expansion, we first write $\hat{U}_{\mathrm{dis}}$ as a Taylor series
\begin{align}
    \hat{U}_{\mathrm{dis}} = \exp\Bigg[\frac{\mathcal{D}(\hat{a}^\dagger-\hat{a})}{\Delta_d}\sum_{k=0}^{\infty} \Bigg(\frac{\chi_L\hat{b}_L^\dagger\hat{b}_L+\chi_R\hat{b}_R^\dagger\hat{b}_R}{\Delta_d}\Bigg)^k
 \Bigg].
\end{align}
Then, to the leading order of $\chi_{L,R}/\Delta_d$, we find
\begin{align}
    \hat{b}_{L(R)}\rightarrow &\,\hat{U}^\dagger_{\mathrm{dis}}\hat{b}_{L(R)}\hat{U}_{\mathrm{dis}}\nonumber\\
    \approx&\,\hat{b}_{L(R)}  +\frac{{\mathcal{D}\chi_{L(R)}}}{\Delta^2_d}(\hat{a}^\dagger-\hat{a})\hat{b}_{L(R)}.
\end{align}
Note that in the displaced frame where the Hamiltonian \eqref{eq:tildeHdisplace} is defined, the resonator frequency is only $\Delta_d$. Therefore, if we apply a drive on the qubits close to $\omega_{L,R}$, the four-wave-mixing interaction is only detuned by $\Delta_d$, which may cause the resonator to leave its displaced vacuum once the coefficient $\mathcal{D}\chi_{L(R)}/\Delta_d^2$ is not sufficiently small.

For the hardware and drive parameters we choose for the $ZZ$ cancellation and the {\footnotesize CR }gates, this four-wave-mixing interaction has not introduced additional gate errors, which we have checked by numerical simulation. [Note that our simulation is based on Eq.~\eqref{eq:dispersive}, where this type of interaction is fully included.] For the experimental implementation of this scheme, one can suppress or mitigate this potential error channel by carefully choosing the detuning $\Delta_d$.

\section{$ZZ$-free architecture for a larger qubit network}
\label{sec:scaleup}

\begin{figure}[t!]
    \centering
    \includegraphics[width = 8.6cm]{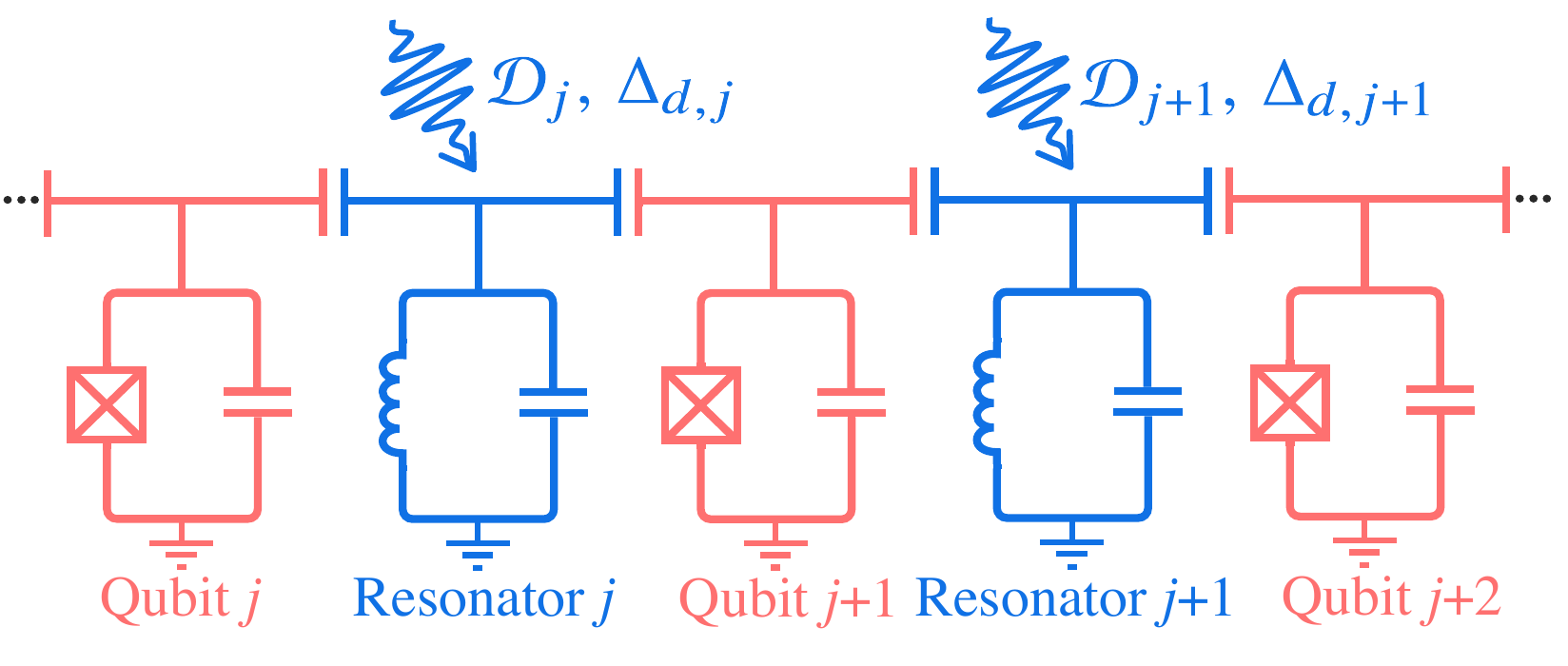}
    \caption{Schematic of a multi-transmon chain connected by cavities. The drive on each resonator can cancel the static $ZZ$ interaction between adjacent qubits, according to Eq.~\eqref{eq:scaleupcancel}. }
    \label{fig:scaleup}
\end{figure}

In this section, we outline a plan to scale up this $ZZ$-free architecture toward a larger qubit network. We consider a transmon-resonator chain that consists of $N$ qubits and $N-1$ cavities. The schematic of such a chain is shown in Fig.~\ref{fig:scaleup}. The adjacent qubit pair indexed by $(j,j+1)$ in this chain are both coupled by the $j$th resonator.  The Hamiltonian of such a chain is given by
\begin{align}
    \hat{H}_\mathrm{chain} = & \,\sum^N_{j=1}\Big(\omega_{q,j} \hat{b}^\dagger_j\hat{b}_j + \frac{\eta_{q,j}}{2}\hat{b}^\dagger_j\hat{b}^\dagger_j\hat{b}_j\hat{b}_j\Big)+\sum_{j=1}^{N-1}\omega_{C,j}\hat{a}^\dagger_j\hat{a}_j\nonumber\\
    &+ \sum^{N-1}_{j=1} \chi'_{j,j+1} \hat{b}^\dagger_{j+1}\hat{b}_{j+1}\hat{b}^\dagger_{j}\hat{b}_{j}\nonumber\\
    &+\sum_{j=1}^{N-1}  \Big(\chi_{j,j+1}\hat{b}^\dagger_{j+1}\hat{b}_{j+1}+\chi_{j,j}\hat{b}^\dagger_{j}\hat{b}_{j}\Big)\hat{a}^\dagger_j\hat{a}_j,
\end{align}
where $\hat{a}_j$ ($\hat{b}_j$) is annihilation operators for the $j$th resonator (qubit), and $\omega_{C,j}$ ($\omega_{q,j}$)  denotes its frequency. Other parameters are similarly defined as in the main text. To cancel the stray $ZZ$ interactions described by terms such as $\chi'_{j,j+1}\hat{b}^\dagger_{j+1}\hat{b}_{j+1}\hat{b}^\dagger_{j}\hat{b}_{j}$, we can set the driving strength $\mathcal{D}_j$ as the solution of the equation
\begin{align}
    \mathcal{D}^2_j\Bigg(&\frac{1}{\Delta_{d,j}-\chi_{j,j}-\chi_{j,j+1}} + \frac{1}{\Delta_{d,j}}\nonumber\\
    &- \frac{1}{\Delta_{d,j}-\chi_{j,j}} - \frac{1}{\Delta_{d,j}-\chi_{j,j+1}}\Bigg)+\chi'_{j,j+1}=0.\label{eq:scaleupcancel} 
\end{align}
Note that there is freedom in choosing the detuning $\Delta_{d,j}$, which provides a way to avoid crosstalks in resonator drives. 

In realistic devices, however, a transmon qubit in such a chain can have small but non-zero $ZZ$ coupling with a distant resonator or qubit. In Appendix \ref{App:scaleup}, we numerically simulate a three-transmon chain as a minimal example to discuss the residual $ZZ$ interactions and potential mitigation solutions.

\section{Conclusion}
\label{sec:conclusion}

In conclusion, we propose a simple architecture where the $ZZ$ coupling between transmon qubits can be dynamically canceled via the {\footnotesize RIP }interaction. The coupler considered here is a high-coherence resonator controlled by a microwave drive, which preserves the high coherence times of the fixed-frequency transmon qubits. The $ZZ$ cancellation enables strong coupling between qubits without introducing stray inter-qubit interaction, which significantly accelerates two-qubit entangling gates. Using realistic parameters, we numerically demonstrate two types of short and high-fidelity entangling gates, namely, cross-resonance {\footnotesize CNOT} gates within 40 ns and adiabatic {\footnotesize CZ} gates within 140 ns. As a showcase, the error of the cross-resonance gate on this architecture can reach below $10^{-4}$ based on our numerical simulation. Besides enabling short gate times and high fidelities, this scheme also allows a relatively larger separation between the qubit frequencies, which can ease the precise requirement of the transmon parameters. We also show that such an architecture is straightforward to scale up.

These advantages make our scheme competitive toward realizing error-correctable quantum computing, especially with the rapid advances in the material improvement of the fixed-frequency transmon qubits. In the future, one can also explore similar $ZZ$-free schemes with other novel superconducting qubits \cite{Gyenis_protected_review}. 

\acknowledgements{This material is based upon work supported by the U.S. Department of Energy, Office of Science, National Quantum Information Science Research Centers, Superconducting Quantum Materials and Systems Center (SQMS) under contract number DE-AC02-07CH11359. We thank Jens Koch and Hanhee Paik for the helpful discussion.}

\appendix

\section{$ZZ$ cancellation for a three-transmon chain}\label{App:scaleup}

\begin{figure}
    \centering
    \includegraphics[width = 8.6cm]{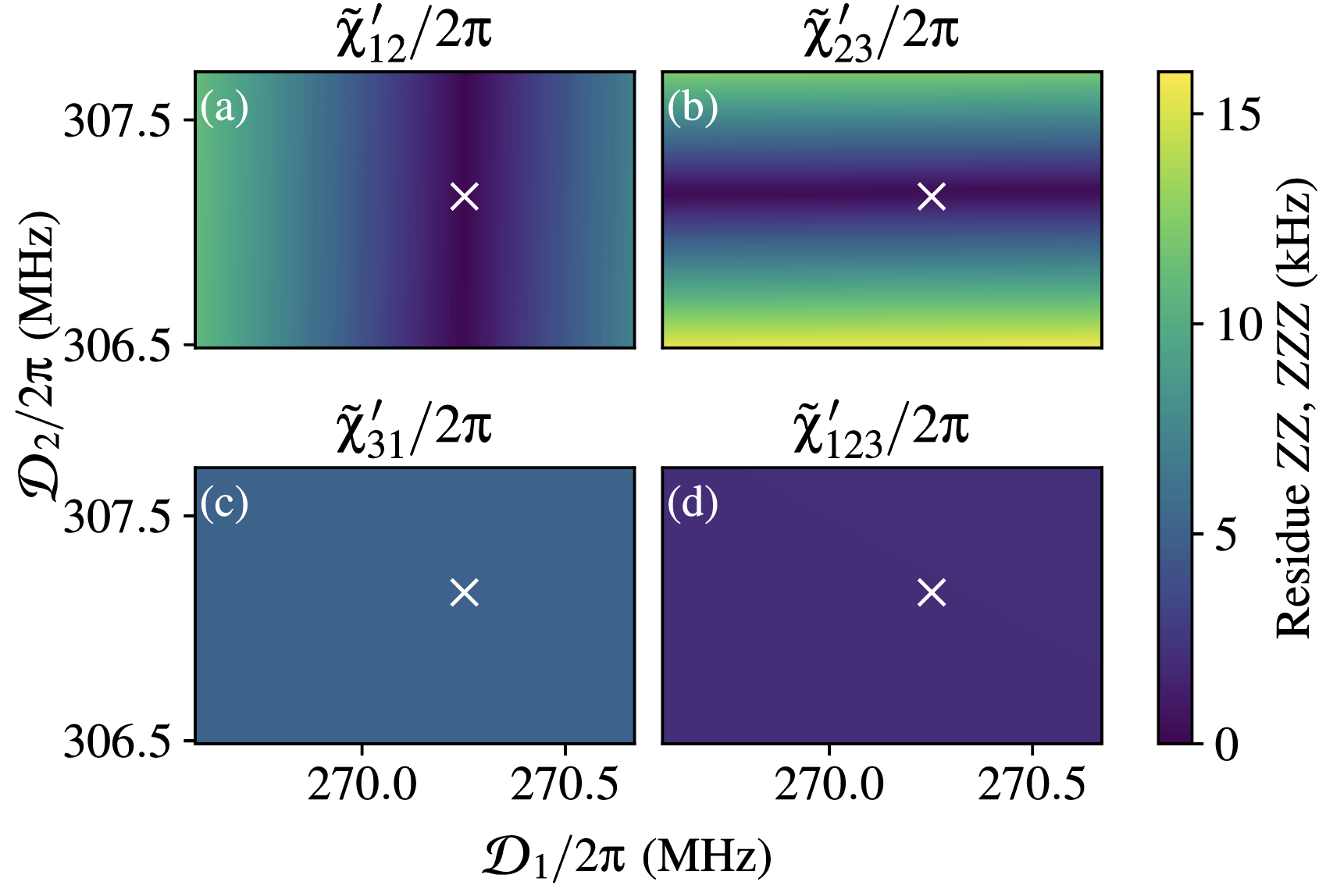}
    \caption{Residue stray couplings in a three-transmon chain after {\footnotesize RIP }$ZZ$ suppression. The frequencies of the three transmons are $\omega_{q,1}/2\pi = 4.24$ GHz, $\omega_{q,2}/2\pi = 4.89$ GHz, and $\omega_{q,3}/2\pi = 5.42$ GHz; their anharmonicies are $\eta_{q,1}/2\pi = -317$ MHz, $\eta_{q,2}/2\pi = -341$ MHz, and $\eta_{q,3}/2\pi = -299$ MHz.  
    The strengths of the dispersive coupling between the transmon qubits and the cavities are: $\chi_{11}/2\pi = -3.66$ MHz, $\chi_{12}/2\pi = -4.05$ kHz, $\chi_{21}/2\pi = -4.96$ MHz, $\chi_{22}/2\pi = -3.61$ MHz, $\chi_{31}/2\pi = -13.8$ kHz, $\chi_{32}/2\pi = -4.44$ MHz. The static $ZZ$ coupling strengths are $\chi'_{12}/2\pi=-2.258$ MHz, $\chi'_{23}/2\pi = -3.45$ MHz and $\chi'_{31}/2\pi=-0.1$ kHz.
    The detunings of the drive on the resonators are both set by $\Delta_{d,1}/2\pi=\Delta_{d,2}/2\pi=100$ MHz.
    }
    \label{fig:ZZ_three_tmon}
\end{figure}

In this appendix, we show the $ZZ$ suppression in a chain of three transmon qubits (labeled as qubit 1, 2 and 3) as an example of the architecture shown in Fig.~\ref{fig:scaleup}. The three transmon qubits are coupled by two resonators (labeled as resonator 1 and 2). We list the relevant parameters of the qubits and resonators in the caption of Fig.~\ref{fig:ZZ_three_tmon}. The magnitudes of the static $ZZ$ coupling between adjacent qubits ($\chi'_{12}$ and $\chi'_{23}$) are a few MHz, which we aim to cancel via the {\footnotesize RIP }interaction. 

By fine-tuning the drive strengths around the theoretical predictions via Eq.~\eqref{eq:scaleupcancel}, we find that the static $ZZ$ interactions can be mostly canceled. In Fig.~\ref{fig:ZZ_three_tmon}, we plot the strengths of the residue $ZZ$ couplings between both the adjacent qubits ($\tilde{\chi}'_{12}$ and $\tilde{\chi}'_{23}$) in Fig.~\ref{fig:ZZ_three_tmon}, both of which vanish for the drive strengths indicated by the white cross. However, due to the dispersive coupling between distant qubits and resonators, the {\footnotesize RIP }drives also lead to two additional interactions: the first is the $ZZ$ interaction between distant qubits, whose strength is denoted by $\tilde{\chi}'_{31}$ [Fig.~\ref{fig:ZZ_three_tmon} (c)]; the second is a $ZZZ$ interaction among the three qubits, whose strength is denoted by $\tilde{\chi}'_{123}$ [Fig.~\ref{fig:ZZ_three_tmon} (d)]. Fortunately, the magnitudes of them are only at kHz level. (At the white cross, we find $\tilde{\chi}'_{31}/2\pi\approx 5.1$ kHz and $\tilde{\chi}'_{123}/2\pi\approx 2.1$ kHz.)

Although the remaining kHz-level $ZZ$ couplings are already weak compared to those in many state-of-the-art two-qubit gates, they can be further mitigated by introducing extra coupler cavities. For example, we can further couple transmon 1 and transmon 3 with an additional resonator and displace the state of that resonator toward the cancellation of all stray interactions.

\section{Alternative approach to deriving the dynamical $ZZ$ coupling}
\label{sizzle_approach}

In this appendix, we describe an alternative method to derive the strength of the dynamical $ZZ$ interaction. The framework introduced here can also be useful for more efficient numerical simulation of our scheme.

Different from the derivation of Eq.~\eqref{eq:dispersive}, here we start from Eq.~\eqref{eq:Hbarefull}, and directly perform a displacement transformation 
$\hat{U}'_{\mathrm{dis}}(t) = \exp[\mathcal{D}(\hat{a}^\dagger-\hat{a})/\Delta_d]\exp(-i\omega^\mathrm{d}_Ct\hat{a}^\dagger\hat{a})$ to adiabatically eliminate the drive term Eq.~\eqref{eq:cavity_drive}. After this step, the displaced Hamiltonian is transformed to
\begin{align}
   \tilde{H}'_{\mathrm{dis}} =&\, \hat{H}_L + \hat{H}_R -\Delta_d\hat{a}^\dagger\hat{a} \nonumber\\
   &+\big(\hat{a}^\dagger e^{i\omega^\mathrm{d}_Ct}\!+\!\hat{a}e^{-i\omega^\mathrm{d}_Ct}\big)\big[\tilde{g}_L(\hat{b}^\dagger_L\!+\!\hat{b}_L)+\tilde{g}_R(\hat{b}^\dagger_R\!+\!\hat{b}_R)\big]\nonumber\\
   &+\frac{2\mathcal{D}}{\Delta_d}\cos \omega^\mathrm{d}_Ct \big[\tilde{g}_L(\hat{b}^\dagger_L\!+\!\hat{b}_L)+\tilde{g}_R(\hat{b}^\dagger_R\!+\!\hat{b}_R)\big].
\end{align}
Above, the displacement drive on the resonator is converted to drives on the two qubits. To further simplify this Hamiltonian, we approximate the transmon Hamiltonian by that of a duffing oscillator, i.e.,
\begin{align}
    \hat{H}_{L(R)} = \bar{\omega}_{L(R)}\hat{b}^\dagger_{L(R)}\hat{b}_{L(R)} + \frac{\eta_{L(R)}}{2}\hat{b}^\dagger_{L(R)}\hat{b}^\dagger_{L(R)}\hat{b}_{L(R)}\hat{b}_{L(R)},
\end{align}
and perform an additional unitary transformation according to $\hat{U}''(t) = \exp[-i\omega^\mathrm{d}_Ct(\hat{b}^\dagger_L\hat{b}_L+\hat{b}_R^\dagger\hat{b}_R)]$. After discarding fast-rotating terms, we arrive at
\begin{align}
    \tilde{H}''_\mathrm{dis} = &\,(\bar{\omega}_{L}-\omega_C^\mathrm{d})\hat{b}^\dagger_{L}\hat{b}_{L} + \frac{\eta_L}{2}\hat{b}_L^\dagger \hat{b}_L^\dagger\hat{b}_L\hat{b}_L \nonumber\\
    &\,+ \tilde{g}_L(\hat{a}^\dagger\hat{b}_L+\hat{a}\hat{b}^\dagger_L)+{g_L\alpha}(\hat{b}^\dagger_L+\hat{b}_L)+ (L\xleftrightarrow{} R)\nonumber\\
    &\,-\Delta_d\hat{a}^\dagger\hat{a}.\label{eq:H''dis}
\end{align}
where we define $\alpha = \mathcal{D}/\Delta_d$.

The transmon displacement terms $\hat{b}^\dagger_{L(R)}+\hat{b}_{L(R)}$ hybridizes the bare transmon states, which eventually leads to the dynamical $ZZ$ interaction \cite{IBM_sizzle}. To derive such hybridization, we consider a third transformation
\begin{align}
    \hat{U}_\mathrm{SW} = \exp\Bigg[&  \hat{b}^\dagger_L\frac{-\tilde{g}_L\alpha}{\bar{\omega}_L-\omega^\mathrm{d}_C+\eta_L\hat{b}_L^\dagger\hat{b}_L} \nonumber\\
    &+\hat{b}^\dagger_R\frac{-\tilde{g}_R\alpha}{\bar{\omega}_R-\omega^\mathrm{d}_C+\eta_R\hat{b}_R^\dagger\hat{b}_R}-\mathrm{H.c.}\Bigg],
\end{align}
which to the leading order of $\tilde{g}_{L(R)}\alpha/[\bar{\omega}_{L(R)}-\omega_C^\mathrm{d}]$ eliminates the transmon displacement terms. Importantly, the transformation described above also modifies the coupling term such as $\tilde{g}_L(\hat{a}^\dagger\hat{b}_L+\hat{a}\hat{b}^\dagger_L)$, which causes the energy shifts of the four transmon states. To derive the transformed qubit-resonator coupling, we first inspect the transformation of $\hat{b}_{L(R)}$. Again, to the leading order, we find
\begin{align}
    \hat{b}_L\rightarrow&\, \hat{U}^\dagger_\mathrm{SW}\hat{b}_L\hat{U}_\mathrm{SW}\nonumber\\
    \approx&\, \hat{b}_L + f(\hat{b}_{L}^\dagger\hat{b}_L) \nonumber\\
    &+ \hat{b}^\dagger_L\hat{b}_L [f(\hat{b}^\dagger_L\hat{b}_L)-f(\hat{b}^\dagger_L\hat{b}_L-1)]\nonumber\\
    &+\hat{b}_L^2[f(b_L^\dagger\hat{b}_L) - f(b_L^\dagger\hat{b}_L+1)],
\end{align}
where we define $f(x) \equiv -\Tilde{g}_L\alpha/[2(\bar{\omega}_L-\omega^\mathrm{d}_L+\eta_L x)]$. 

In the limit of $|\omega_{L(R)}-\omega_C^\mathrm{d}|\gg \eta_{L(R)}$, we can further approximate this transformation to the leading order of $\eta_{L(R)}/|\omega_{L(R)}-\omega_C^\mathrm{d}|$ by,
\begin{align}
    \hat{b}_L\rightarrow&\, \hat{U}^\dagger_\mathrm{SW}\hat{b}_L\hat{U}_\mathrm{SW}\nonumber\\
    \approx&\, \hat{b}_L - \frac{\tilde{g}_L\alpha}{\big(\bar{\omega}_L -\omega^\mathrm{d}_C\big)}\nonumber\\
    &+  \frac{\tilde{g}_L\alpha\eta_L}{\big(\bar{\omega}_L -\omega^\mathrm{d}_C\big)^2}(2\hat{b}^\dagger_L\hat{b}_L-\hat{b}_L^2). \label{eq:blexp}
\end{align}
The useful term in this expansion for the generation of $ZZ$ coupling is $\hat{b}^\dagger_L\hat{b}_L$, while the squeezing term $\hat{b}^2_L$ is less important. If the two qubits are coupled by direct exchange interaction $J(\hat{b}^\dagger_L\hat{b}_R+\hat{b}^\dagger_R\hat{b}_L)$, inserting $\hat{b}^\dagger_L\hat{b}_L$ and similarly $\hat{b}^\dagger_R\hat{b}_R$ into the previous $\hat{b}_L$ and $\hat{b}_R$ yields the ``siZZle" dynamical $ZZ$ coupling \cite{IBM_sizzle}. For our case, the qubit-resonator coupling $\tilde{g}_{L(R)}[\hat{a}\hat{b}^\dagger_{L(R)}+\hat{a}^\dagger\hat{b}_{L(R)}]$ instead yields a term
\begin{align}
\alpha(\chi_L\hat{b}^\dagger_L\hat{b}_L+\chi_R\hat{b}^\dagger_R\hat{b}_R)(\hat{a}^\dagger+\hat{a}),
\end{align}
where we have used the approximation $\chi_{L(R)}\approx 2\tilde{g}^2_{L(R)}\eta_{L(R)}/[\bar{\omega}_{L(R)}-\omega^\mathrm{d}_C]^2$ ($\omega^\mathrm{d}_C$ is set to be detuned from but close to $\omega_C$) \footnote{More accurately, this coefficient is derived as $\chi_{L(R)} = 2\tilde{g}^2_{L(R)}\eta_{L(R)}/[(\bar{\omega}_{L(R)}-\bar{\omega}_C)(\bar{\omega}_{L(R)}-\bar{\omega}_C+\eta_{L(R)})]$ \cite{Blais_cqed_review}.}. Finally, we only need to integrate out the resonator degree of freedom, which can be done by a Born-Oppenheimer approximation \cite{Dvir_BO,Huang_dissertation}. This step results in a two-qubit interaction term
\begin{align}
    -\frac{\alpha^2}{\Delta_d}(\chi_L\hat{b}_L^\dagger\hat{b}_L+\chi_R\hat{b}^\dagger_R\hat{b}_R)^2,
\end{align}
which reproduces Eq.~\eqref{eq:ZZapproximation}.

Besides cross-checking the strength of the dynamical $ZZ$ interaction, the framework detailed in Eq.~\eqref{eq:H''dis} can also be useful for faster numerical simulation. In this frame, the resonator does not possess a large number of photons, which can reduce the dimension of the Hilbert space and further the numerical cost.

%


\begin{thebibliography}{63}%
\makeatletter
\providecommand \@ifxundefined [1]{%
 \@ifx{#1\undefined}
}%
\providecommand \@ifnum [1]{%
 \ifnum #1\expandafter \@firstoftwo
 \else \expandafter \@secondoftwo
 \fi
}%
\providecommand \@ifx [1]{%
 \ifx #1\expandafter \@firstoftwo
 \else \expandafter \@secondoftwo
 \fi
}%
\providecommand \natexlab [1]{#1}%
\providecommand \enquote  [1]{``#1''}%
\providecommand \bibnamefont  [1]{#1}%
\providecommand \bibfnamefont [1]{#1}%
\providecommand \citenamefont [1]{#1}%
\providecommand \href@noop [0]{\@secondoftwo}%
\providecommand \href [0]{\begingroup \@sanitize@url \@href}%
\providecommand \@href[1]{\@@startlink{#1}\@@href}%
\providecommand \@@href[1]{\endgroup#1\@@endlink}%
\providecommand \@sanitize@url [0]{\catcode `\\12\catcode `\$12\catcode
  `\&12\catcode `\#12\catcode `\^12\catcode `\_12\catcode `\%12\relax}%
\providecommand \@@startlink[1]{}%
\providecommand \@@endlink[0]{}%
\providecommand \url  [0]{\begingroup\@sanitize@url \@url }%
\providecommand \@url [1]{\endgroup\@href {#1}{\urlprefix }}%
\providecommand \urlprefix  [0]{URL }%
\providecommand \Eprint [0]{\href }%
\providecommand \doibase [0]{https://doi.org/}%
\providecommand \selectlanguage [0]{\@gobble}%
\providecommand \bibinfo  [0]{\@secondoftwo}%
\providecommand \bibfield  [0]{\@secondoftwo}%
\providecommand \translation [1]{[#1]}%
\providecommand \BibitemOpen [0]{}%
\providecommand \bibitemStop [0]{}%
\providecommand \bibitemNoStop [0]{.\EOS\space}%
\providecommand \EOS [0]{\spacefactor3000\relax}%
\providecommand \BibitemShut  [1]{\csname bibitem#1\endcsname}%
\let\auto@bib@innerbib\@empty
\bibitem [{\citenamefont {Blais}\ \emph {et~al.}(2021)\citenamefont {Blais},
  \citenamefont {Grimsmo}, \citenamefont {Girvin},\ and\ \citenamefont
  {Wallraff}}]{Blais_cqed_review}%
  \BibitemOpen
  \bibfield  {author} {\bibinfo {author} {\bibfnamefont {A.}~\bibnamefont
  {Blais}}, \bibinfo {author} {\bibfnamefont {A.~L.}\ \bibnamefont {Grimsmo}},
  \bibinfo {author} {\bibfnamefont {S.~M.}\ \bibnamefont {Girvin}},\ and\
  \bibinfo {author} {\bibfnamefont {A.}~\bibnamefont {Wallraff}},\ }\bibfield
  {title} {\bibinfo {title} {\textit{Circuit Quantum Electrodynamics}},\ }\href
  {https://doi.org/10.1103/RevModPhys.93.025005} {\bibfield  {journal}
  {\bibinfo  {journal} {Rev. Mod. Phys.}\ }\textbf {\bibinfo {volume} {93}},\
  \bibinfo {pages} {025005} (\bibinfo {year} {2021})}\BibitemShut {NoStop}%
\bibitem [{\citenamefont {Kjaergaard}\ \emph {et~al.}(2020)\citenamefont
  {Kjaergaard}, \citenamefont {Schwartz}, \citenamefont {Braumüller},
  \citenamefont {Krantz}, \citenamefont {Wang}, \citenamefont {Gustavsson},\
  and\ \citenamefont {Oliver}}]{Oliver_scqubits_review}%
  \BibitemOpen
  \bibfield  {author} {\bibinfo {author} {\bibfnamefont {M.}~\bibnamefont
  {Kjaergaard}}, \bibinfo {author} {\bibfnamefont {M.~E.}\ \bibnamefont
  {Schwartz}}, \bibinfo {author} {\bibfnamefont {J.}~\bibnamefont
  {Braumüller}}, \bibinfo {author} {\bibfnamefont {P.}~\bibnamefont {Krantz}},
  \bibinfo {author} {\bibfnamefont {J.~I.-J.}\ \bibnamefont {Wang}}, \bibinfo
  {author} {\bibfnamefont {S.}~\bibnamefont {Gustavsson}},\ and\ \bibinfo
  {author} {\bibfnamefont {W.~D.}\ \bibnamefont {Oliver}},\ }\bibfield  {title}
  {\bibinfo {title} {\textit{Superconducting Qubits: Current State of Play}},\
  }\href {https://doi.org/10.1146/annurev-conmatphys-031119-050605} {\bibfield
  {journal} {\bibinfo  {journal} {Annu. Rev. Condens. Matter Phys.}\ }\textbf
  {\bibinfo {volume} {11}},\ \bibinfo {pages} {369} (\bibinfo {year}
  {2020})}\BibitemShut {NoStop}%
\bibitem [{\citenamefont {Yan}\ \emph {et~al.}(2018)\citenamefont {Yan},
  \citenamefont {Krantz}, \citenamefont {Sung}, \citenamefont {Kjaergaard},
  \citenamefont {Campbell}, \citenamefont {Orlando}, \citenamefont
  {Gustavsson},\ and\ \citenamefont {Oliver}}]{Oliver_tunable_coupling}%
  \BibitemOpen
  \bibfield  {author} {\bibinfo {author} {\bibfnamefont {F.}~\bibnamefont
  {Yan}}, \bibinfo {author} {\bibfnamefont {P.}~\bibnamefont {Krantz}},
  \bibinfo {author} {\bibfnamefont {Y.}~\bibnamefont {Sung}}, \bibinfo {author}
  {\bibfnamefont {M.}~\bibnamefont {Kjaergaard}}, \bibinfo {author}
  {\bibfnamefont {D.~L.}\ \bibnamefont {Campbell}}, \bibinfo {author}
  {\bibfnamefont {T.~P.}\ \bibnamefont {Orlando}}, \bibinfo {author}
  {\bibfnamefont {S.}~\bibnamefont {Gustavsson}},\ and\ \bibinfo {author}
  {\bibfnamefont {W.~D.}\ \bibnamefont {Oliver}},\ }\bibfield  {title}
  {\bibinfo {title} {\textit{Tunable Coupling Scheme for Implementing
  High-Fidelity Two-Qubit Gates}},\ }\href
  {https://doi.org/10.1103/PhysRevApplied.10.054062} {\bibfield  {journal}
  {\bibinfo  {journal} {Phys. Rev. Appl.}\ }\textbf {\bibinfo {volume} {10}},\
  \bibinfo {pages} {054062} (\bibinfo {year} {2018})}\BibitemShut {NoStop}%
\bibitem [{\citenamefont {Mundada}\ \emph {et~al.}(2019)\citenamefont
  {Mundada}, \citenamefont {Zhang}, \citenamefont {Hazard},\ and\ \citenamefont
  {Houck}}]{Houck_ZZ_free}%
  \BibitemOpen
  \bibfield  {author} {\bibinfo {author} {\bibfnamefont {P.}~\bibnamefont
  {Mundada}}, \bibinfo {author} {\bibfnamefont {G.}~\bibnamefont {Zhang}},
  \bibinfo {author} {\bibfnamefont {T.}~\bibnamefont {Hazard}},\ and\ \bibinfo
  {author} {\bibfnamefont {A.}~\bibnamefont {Houck}},\ }\bibfield  {title}
  {\bibinfo {title} {\textit{Suppression of Qubit Crosstalk in a Tunable
  Coupling Superconducting Circuit}},\ }\href
  {https://doi.org/10.1103/PhysRevApplied.12.054023} {\bibfield  {journal}
  {\bibinfo  {journal} {Phys. Rev. Appl.}\ }\textbf {\bibinfo {volume} {12}},\
  \bibinfo {pages} {054023} (\bibinfo {year} {2019})}\BibitemShut {NoStop}%
\bibitem [{\citenamefont {Srinivasan}\ \emph {et~al.}(2011)\citenamefont
  {Srinivasan}, \citenamefont {Hoffman}, \citenamefont {Gambetta},\ and\
  \citenamefont {Houck}}]{Houck_tunable_coupling}%
  \BibitemOpen
  \bibfield  {author} {\bibinfo {author} {\bibfnamefont {S.~J.}\ \bibnamefont
  {Srinivasan}}, \bibinfo {author} {\bibfnamefont {A.~J.}\ \bibnamefont
  {Hoffman}}, \bibinfo {author} {\bibfnamefont {J.~M.}\ \bibnamefont
  {Gambetta}},\ and\ \bibinfo {author} {\bibfnamefont {A.~A.}\ \bibnamefont
  {Houck}},\ }\bibfield  {title} {\bibinfo {title} {\textit{Tunable Coupling in
  Circuit Quantum Electrodynamics Using a Superconducting Charge Qubit with a
  $V$-Shaped Energy Level Diagram}},\ }\href
  {https://doi.org/10.1103/PhysRevLett.106.083601} {\bibfield  {journal}
  {\bibinfo  {journal} {Phys. Rev. Lett.}\ }\textbf {\bibinfo {volume} {106}},\
  \bibinfo {pages} {083601} (\bibinfo {year} {2011})}\BibitemShut {NoStop}%
\bibitem [{\citenamefont {Sung}\ \emph {et~al.}(2021)\citenamefont {Sung},
  \citenamefont {Ding}, \citenamefont {Braum\"uller}, \citenamefont
  {Veps\"al\"ainen}, \citenamefont {Kannan}, \citenamefont {Kjaergaard},
  \citenamefont {Greene}, \citenamefont {Samach}, \citenamefont {McNally},
  \citenamefont {Kim}, \citenamefont {Melville}, \citenamefont {Niedzielski},
  \citenamefont {Schwartz}, \citenamefont {Yoder}, \citenamefont {Orlando},
  \citenamefont {Gustavsson},\ and\ \citenamefont
  {Oliver}}]{Oliver_two_qubit_gate}%
  \BibitemOpen
  \bibfield  {author} {\bibinfo {author} {\bibfnamefont {Y.}~\bibnamefont
  {Sung}}, \bibinfo {author} {\bibfnamefont {L.}~\bibnamefont {Ding}}, \bibinfo
  {author} {\bibfnamefont {J.}~\bibnamefont {Braum\"uller}}, \bibinfo {author}
  {\bibfnamefont {A.}~\bibnamefont {Veps\"al\"ainen}}, \bibinfo {author}
  {\bibfnamefont {B.}~\bibnamefont {Kannan}}, \bibinfo {author} {\bibfnamefont
  {M.}~\bibnamefont {Kjaergaard}}, \bibinfo {author} {\bibfnamefont
  {A.}~\bibnamefont {Greene}}, \bibinfo {author} {\bibfnamefont {G.~O.}\
  \bibnamefont {Samach}}, \bibinfo {author} {\bibfnamefont {C.}~\bibnamefont
  {McNally}}, \bibinfo {author} {\bibfnamefont {D.}~\bibnamefont {Kim}},
  \bibinfo {author} {\bibfnamefont {A.}~\bibnamefont {Melville}}, \bibinfo
  {author} {\bibfnamefont {B.~M.}\ \bibnamefont {Niedzielski}}, \bibinfo
  {author} {\bibfnamefont {M.~E.}\ \bibnamefont {Schwartz}}, \bibinfo {author}
  {\bibfnamefont {J.~L.}\ \bibnamefont {Yoder}}, \bibinfo {author}
  {\bibfnamefont {T.~P.}\ \bibnamefont {Orlando}}, \bibinfo {author}
  {\bibfnamefont {S.}~\bibnamefont {Gustavsson}},\ and\ \bibinfo {author}
  {\bibfnamefont {W.~D.}\ \bibnamefont {Oliver}},\ }\bibfield  {title}
  {\bibinfo {title} {\textit{Realization of High-Fidelity CZ and $ZZ$-Free
  iSWAP Gates with a Tunable Coupler}},\ }\href
  {https://doi.org/10.1103/PhysRevX.11.021058} {\bibfield  {journal} {\bibinfo
  {journal} {Phys. Rev. X}\ }\textbf {\bibinfo {volume} {11}},\ \bibinfo
  {pages} {021058} (\bibinfo {year} {2021})}\BibitemShut {NoStop}%
\bibitem [{\citenamefont {Xu}\ \emph {et~al.}(2020)\citenamefont {Xu},
  \citenamefont {Chu}, \citenamefont {Yuan}, \citenamefont {Qiu}, \citenamefont
  {Zhou}, \citenamefont {Zhang}, \citenamefont {Tan}, \citenamefont {Yu},
  \citenamefont {Liu}, \citenamefont {Li}, \citenamefont {Yan},\ and\
  \citenamefont {Yu}}]{Yu_two_qubit_gates}%
  \BibitemOpen
  \bibfield  {author} {\bibinfo {author} {\bibfnamefont {Y.}~\bibnamefont
  {Xu}}, \bibinfo {author} {\bibfnamefont {J.}~\bibnamefont {Chu}}, \bibinfo
  {author} {\bibfnamefont {J.}~\bibnamefont {Yuan}}, \bibinfo {author}
  {\bibfnamefont {J.}~\bibnamefont {Qiu}}, \bibinfo {author} {\bibfnamefont
  {Y.}~\bibnamefont {Zhou}}, \bibinfo {author} {\bibfnamefont {L.}~\bibnamefont
  {Zhang}}, \bibinfo {author} {\bibfnamefont {X.}~\bibnamefont {Tan}}, \bibinfo
  {author} {\bibfnamefont {Y.}~\bibnamefont {Yu}}, \bibinfo {author}
  {\bibfnamefont {S.}~\bibnamefont {Liu}}, \bibinfo {author} {\bibfnamefont
  {J.}~\bibnamefont {Li}}, \bibinfo {author} {\bibfnamefont {F.}~\bibnamefont
  {Yan}},\ and\ \bibinfo {author} {\bibfnamefont {D.}~\bibnamefont {Yu}},\
  }\bibfield  {title} {\bibinfo {title} {\textit{High-Fidelity,
  High-Scalability Two-Qubit Gate Scheme for Superconducting Qubits}},\ }\href
  {https://doi.org/10.1103/PhysRevLett.125.240503} {\bibfield  {journal}
  {\bibinfo  {journal} {Phys. Rev. Lett.}\ }\textbf {\bibinfo {volume} {125}},\
  \bibinfo {pages} {240503} (\bibinfo {year} {2020})}\BibitemShut {NoStop}%
\bibitem [{\citenamefont {Ku}\ \emph {et~al.}(2020)\citenamefont {Ku},
  \citenamefont {Xu}, \citenamefont {Brink}, \citenamefont {McKay},
  \citenamefont {Hertzberg}, \citenamefont {Ansari},\ and\ \citenamefont
  {Plourde}}]{Plourde_CR_tunable_coupler}%
  \BibitemOpen
  \bibfield  {author} {\bibinfo {author} {\bibfnamefont {J.}~\bibnamefont
  {Ku}}, \bibinfo {author} {\bibfnamefont {X.}~\bibnamefont {Xu}}, \bibinfo
  {author} {\bibfnamefont {M.}~\bibnamefont {Brink}}, \bibinfo {author}
  {\bibfnamefont {D.~C.}\ \bibnamefont {McKay}}, \bibinfo {author}
  {\bibfnamefont {J.~B.}\ \bibnamefont {Hertzberg}}, \bibinfo {author}
  {\bibfnamefont {M.~H.}\ \bibnamefont {Ansari}},\ and\ \bibinfo {author}
  {\bibfnamefont {B.~L.~T.}\ \bibnamefont {Plourde}},\ }\bibfield  {title}
  {\bibinfo {title} {\textit{Suppression of Unwanted $ZZ$ Interactions in a
  Hybrid Two-Qubit System}},\ }\href
  {https://doi.org/10.1103/PhysRevLett.125.200504} {\bibfield  {journal}
  {\bibinfo  {journal} {Phys. Rev. Lett.}\ }\textbf {\bibinfo {volume} {125}},\
  \bibinfo {pages} {200504} (\bibinfo {year} {2020})}\BibitemShut {NoStop}%
\bibitem [{\citenamefont {Xu}\ and\ \citenamefont
  {Ansari}(2023)}]{Xu_weakly_tunable}%
  \BibitemOpen
  \bibfield  {author} {\bibinfo {author} {\bibfnamefont {X.}~\bibnamefont
  {Xu}}\ and\ \bibinfo {author} {\bibfnamefont {M.}~\bibnamefont {Ansari}},\
  }\bibfield  {title} {\bibinfo {title} {\textit{Parasitic-Free Gate: An
  Error-Protected Cross-Resonance Switch in Weakly Tunable Architectures}},\
  }\href {https://doi.org/10.1103/PhysRevApplied.19.024057} {\bibfield
  {journal} {\bibinfo  {journal} {Phys. Rev. Appl.}\ }\textbf {\bibinfo
  {volume} {19}},\ \bibinfo {pages} {024057} (\bibinfo {year}
  {2023})}\BibitemShut {NoStop}%
\bibitem [{\citenamefont {Sete}\ \emph {et~al.}(2021)\citenamefont {Sete},
  \citenamefont {Chen}, \citenamefont {Manenti}, \citenamefont {Kulshreshtha},\
  and\ \citenamefont {Poletto}}]{Rigetti_tunable_coupler}%
  \BibitemOpen
  \bibfield  {author} {\bibinfo {author} {\bibfnamefont {E.~A.}\ \bibnamefont
  {Sete}}, \bibinfo {author} {\bibfnamefont {A.~Q.}\ \bibnamefont {Chen}},
  \bibinfo {author} {\bibfnamefont {R.}~\bibnamefont {Manenti}}, \bibinfo
  {author} {\bibfnamefont {S.}~\bibnamefont {Kulshreshtha}},\ and\ \bibinfo
  {author} {\bibfnamefont {S.}~\bibnamefont {Poletto}},\ }\bibfield  {title}
  {\bibinfo {title} {\textit{Floating Tunable Coupler for Scalable Quantum
  Computing Architectures}},\ }\href
  {https://doi.org/10.1103/PhysRevApplied.15.064063} {\bibfield  {journal}
  {\bibinfo  {journal} {Phys. Rev. Appl.}\ }\textbf {\bibinfo {volume} {15}},\
  \bibinfo {pages} {064063} (\bibinfo {year} {2021})}\BibitemShut {NoStop}%
\bibitem [{\citenamefont {Li}\ \emph {et~al.}(2020)\citenamefont {Li},
  \citenamefont {Cai}, \citenamefont {Yan}, \citenamefont {Wang}, \citenamefont
  {Pan}, \citenamefont {Ma}, \citenamefont {Cai}, \citenamefont {Han},
  \citenamefont {Hua}, \citenamefont {Han}, \citenamefont {Wu}, \citenamefont
  {Zhang}, \citenamefont {Wang}, \citenamefont {Song}, \citenamefont {Duan},\
  and\ \citenamefont {Sun}}]{Sun_tunable_coupler}%
  \BibitemOpen
  \bibfield  {author} {\bibinfo {author} {\bibfnamefont {X.}~\bibnamefont
  {Li}}, \bibinfo {author} {\bibfnamefont {T.}~\bibnamefont {Cai}}, \bibinfo
  {author} {\bibfnamefont {H.}~\bibnamefont {Yan}}, \bibinfo {author}
  {\bibfnamefont {Z.}~\bibnamefont {Wang}}, \bibinfo {author} {\bibfnamefont
  {X.}~\bibnamefont {Pan}}, \bibinfo {author} {\bibfnamefont {Y.}~\bibnamefont
  {Ma}}, \bibinfo {author} {\bibfnamefont {W.}~\bibnamefont {Cai}}, \bibinfo
  {author} {\bibfnamefont {J.}~\bibnamefont {Han}}, \bibinfo {author}
  {\bibfnamefont {Z.}~\bibnamefont {Hua}}, \bibinfo {author} {\bibfnamefont
  {X.}~\bibnamefont {Han}}, \bibinfo {author} {\bibfnamefont {Y.}~\bibnamefont
  {Wu}}, \bibinfo {author} {\bibfnamefont {H.}~\bibnamefont {Zhang}}, \bibinfo
  {author} {\bibfnamefont {H.}~\bibnamefont {Wang}}, \bibinfo {author}
  {\bibfnamefont {Y.}~\bibnamefont {Song}}, \bibinfo {author} {\bibfnamefont
  {L.}~\bibnamefont {Duan}},\ and\ \bibinfo {author} {\bibfnamefont
  {L.}~\bibnamefont {Sun}},\ }\bibfield  {title} {\bibinfo {title}
  {\textit{Tunable Coupler for Realizing a Controlled-Phase Gate with
  Dynamically Decoupled Regime in a Superconducting Circuit}},\ }\href
  {https://doi.org/10.1103/PhysRevApplied.14.024070} {\bibfield  {journal}
  {\bibinfo  {journal} {Phys. Rev. Appl.}\ }\textbf {\bibinfo {volume} {14}},\
  \bibinfo {pages} {024070} (\bibinfo {year} {2020})}\BibitemShut {NoStop}%
\bibitem [{\citenamefont {Stehlik}\ \emph {et~al.}(2021)\citenamefont
  {Stehlik}, \citenamefont {Zajac}, \citenamefont {Underwood}, \citenamefont
  {Phung}, \citenamefont {Blair}, \citenamefont {Carnevale}, \citenamefont
  {Klaus}, \citenamefont {Keefe}, \citenamefont {Carniol}, \citenamefont
  {Kumph}, \citenamefont {Steffen},\ and\ \citenamefont
  {Dial}}]{IBM_tunable_coupling}%
  \BibitemOpen
  \bibfield  {author} {\bibinfo {author} {\bibfnamefont {J.}~\bibnamefont
  {Stehlik}}, \bibinfo {author} {\bibfnamefont {D.~M.}\ \bibnamefont {Zajac}},
  \bibinfo {author} {\bibfnamefont {D.~L.}\ \bibnamefont {Underwood}}, \bibinfo
  {author} {\bibfnamefont {T.}~\bibnamefont {Phung}}, \bibinfo {author}
  {\bibfnamefont {J.}~\bibnamefont {Blair}}, \bibinfo {author} {\bibfnamefont
  {S.}~\bibnamefont {Carnevale}}, \bibinfo {author} {\bibfnamefont
  {D.}~\bibnamefont {Klaus}}, \bibinfo {author} {\bibfnamefont {G.~A.}\
  \bibnamefont {Keefe}}, \bibinfo {author} {\bibfnamefont {A.}~\bibnamefont
  {Carniol}}, \bibinfo {author} {\bibfnamefont {M.}~\bibnamefont {Kumph}},
  \bibinfo {author} {\bibfnamefont {M.}~\bibnamefont {Steffen}},\ and\ \bibinfo
  {author} {\bibfnamefont {O.~E.}\ \bibnamefont {Dial}},\ }\bibfield  {title}
  {\bibinfo {title} {\textit{Tunable Coupling Architecture for Fixed-Frequency
  Transmon Superconducting Qubits}},\ }\href
  {https://doi.org/10.1103/PhysRevLett.127.080505} {\bibfield  {journal}
  {\bibinfo  {journal} {Phys. Rev. Lett.}\ }\textbf {\bibinfo {volume} {127}},\
  \bibinfo {pages} {080505} (\bibinfo {year} {2021})}\BibitemShut {NoStop}%
\bibitem [{\citenamefont {Koch}\ \emph {et~al.}(2007)\citenamefont {Koch},
  \citenamefont {Yu}, \citenamefont {Gambetta}, \citenamefont {Houck},
  \citenamefont {Schuster}, \citenamefont {Majer}, \citenamefont {Blais},
  \citenamefont {Devoret}, \citenamefont {Girvin},\ and\ \citenamefont
  {Schoelkopf}}]{Koch_transmon_theory}%
  \BibitemOpen
  \bibfield  {author} {\bibinfo {author} {\bibfnamefont {J.}~\bibnamefont
  {Koch}}, \bibinfo {author} {\bibfnamefont {T.~M.}\ \bibnamefont {Yu}},
  \bibinfo {author} {\bibfnamefont {J.}~\bibnamefont {Gambetta}}, \bibinfo
  {author} {\bibfnamefont {A.~A.}\ \bibnamefont {Houck}}, \bibinfo {author}
  {\bibfnamefont {D.~I.}\ \bibnamefont {Schuster}}, \bibinfo {author}
  {\bibfnamefont {J.}~\bibnamefont {Majer}}, \bibinfo {author} {\bibfnamefont
  {A.}~\bibnamefont {Blais}}, \bibinfo {author} {\bibfnamefont {M.~H.}\
  \bibnamefont {Devoret}}, \bibinfo {author} {\bibfnamefont {S.~M.}\
  \bibnamefont {Girvin}},\ and\ \bibinfo {author} {\bibfnamefont {R.~J.}\
  \bibnamefont {Schoelkopf}},\ }\bibfield  {title} {\bibinfo {title}
  {\textit{Charge-Insensitive Qubit Design Derived from the Cooper Pair Box}},\
  }\href {https://doi.org/10.1103/PhysRevA.76.042319} {\bibfield  {journal}
  {\bibinfo  {journal} {Phys. Rev. A}\ }\textbf {\bibinfo {volume} {76}},\
  \bibinfo {pages} {042319} (\bibinfo {year} {2007})}\BibitemShut {NoStop}%
\bibitem [{\citenamefont {Reagor}\ \emph {et~al.}(2018)\citenamefont {Reagor},
  \citenamefont {Osborn}, \citenamefont {Tezak}, \citenamefont {Staley},
  \citenamefont {Prawiroatmodjo}, \citenamefont {Scheer}, \citenamefont
  {Alidoust}, \citenamefont {Sete}, \citenamefont {Didier}, \citenamefont
  {da~Silva}, \citenamefont {Acala}, \citenamefont {Angeles}, \citenamefont
  {Bestwick}, \citenamefont {Block}, \citenamefont {Bloom}, \citenamefont
  {Bradley}, \citenamefont {Bui}, \citenamefont {Caldwell}, \citenamefont
  {Capelluto}, \citenamefont {Chilcott}, \citenamefont {Cordova}, \citenamefont
  {Crossman}, \citenamefont {Curtis}, \citenamefont {Deshpande}, \citenamefont
  {Bouayadi}, \citenamefont {Girshovich}, \citenamefont {Hong}, \citenamefont
  {Hudson}, \citenamefont {Karalekas}, \citenamefont {Kuang}, \citenamefont
  {Lenihan}, \citenamefont {Manenti}, \citenamefont {Manning}, \citenamefont
  {Marshall}, \citenamefont {Mohan}, \citenamefont {O’Brien}, \citenamefont
  {Otterbach}, \citenamefont {Papageorge}, \citenamefont {Paquette},
  \citenamefont {Pelstring}, \citenamefont {Polloreno}, \citenamefont {Rawat},
  \citenamefont {Ryan}, \citenamefont {Renzas}, \citenamefont {Rubin},
  \citenamefont {Russel}, \citenamefont {Rust}, \citenamefont {Scarabelli},
  \citenamefont {Selvanayagam}, \citenamefont {Sinclair}, \citenamefont
  {Smith}, \citenamefont {Suska}, \citenamefont {To}, \citenamefont
  {Vahidpour}, \citenamefont {Vodrahalli}, \citenamefont {Whyland},
  \citenamefont {Yadav}, \citenamefont {Zeng},\ and\ \citenamefont
  {Rigetti}}]{rigetti_qpu}%
  \BibitemOpen
  \bibfield  {author} {\bibinfo {author} {\bibfnamefont {M.}~\bibnamefont
  {Reagor}}, \bibinfo {author} {\bibfnamefont {C.~B.}\ \bibnamefont {Osborn}},
  \bibinfo {author} {\bibfnamefont {N.}~\bibnamefont {Tezak}}, \bibinfo
  {author} {\bibfnamefont {A.}~\bibnamefont {Staley}}, \bibinfo {author}
  {\bibfnamefont {G.}~\bibnamefont {Prawiroatmodjo}}, \bibinfo {author}
  {\bibfnamefont {M.}~\bibnamefont {Scheer}}, \bibinfo {author} {\bibfnamefont
  {N.}~\bibnamefont {Alidoust}}, \bibinfo {author} {\bibfnamefont {E.~A.}\
  \bibnamefont {Sete}}, \bibinfo {author} {\bibfnamefont {N.}~\bibnamefont
  {Didier}}, \bibinfo {author} {\bibfnamefont {M.~P.}\ \bibnamefont
  {da~Silva}}, \bibinfo {author} {\bibfnamefont {E.}~\bibnamefont {Acala}},
  \bibinfo {author} {\bibfnamefont {J.}~\bibnamefont {Angeles}}, \bibinfo
  {author} {\bibfnamefont {A.}~\bibnamefont {Bestwick}}, \bibinfo {author}
  {\bibfnamefont {M.}~\bibnamefont {Block}}, \bibinfo {author} {\bibfnamefont
  {B.}~\bibnamefont {Bloom}}, \bibinfo {author} {\bibfnamefont
  {A.}~\bibnamefont {Bradley}}, \bibinfo {author} {\bibfnamefont
  {C.}~\bibnamefont {Bui}}, \bibinfo {author} {\bibfnamefont {S.}~\bibnamefont
  {Caldwell}}, \bibinfo {author} {\bibfnamefont {L.}~\bibnamefont {Capelluto}},
  \bibinfo {author} {\bibfnamefont {R.}~\bibnamefont {Chilcott}}, \bibinfo
  {author} {\bibfnamefont {J.}~\bibnamefont {Cordova}}, \bibinfo {author}
  {\bibfnamefont {G.}~\bibnamefont {Crossman}}, \bibinfo {author}
  {\bibfnamefont {M.}~\bibnamefont {Curtis}}, \bibinfo {author} {\bibfnamefont
  {S.}~\bibnamefont {Deshpande}}, \bibinfo {author} {\bibfnamefont {T.~E.}\
  \bibnamefont {Bouayadi}}, \bibinfo {author} {\bibfnamefont {D.}~\bibnamefont
  {Girshovich}}, \bibinfo {author} {\bibfnamefont {S.}~\bibnamefont {Hong}},
  \bibinfo {author} {\bibfnamefont {A.}~\bibnamefont {Hudson}}, \bibinfo
  {author} {\bibfnamefont {P.}~\bibnamefont {Karalekas}}, \bibinfo {author}
  {\bibfnamefont {K.}~\bibnamefont {Kuang}}, \bibinfo {author} {\bibfnamefont
  {M.}~\bibnamefont {Lenihan}}, \bibinfo {author} {\bibfnamefont
  {R.}~\bibnamefont {Manenti}}, \bibinfo {author} {\bibfnamefont
  {T.}~\bibnamefont {Manning}}, \bibinfo {author} {\bibfnamefont
  {J.}~\bibnamefont {Marshall}}, \bibinfo {author} {\bibfnamefont
  {Y.}~\bibnamefont {Mohan}}, \bibinfo {author} {\bibfnamefont
  {W.}~\bibnamefont {O’Brien}}, \bibinfo {author} {\bibfnamefont
  {J.}~\bibnamefont {Otterbach}}, \bibinfo {author} {\bibfnamefont
  {A.}~\bibnamefont {Papageorge}}, \bibinfo {author} {\bibfnamefont {J.-P.}\
  \bibnamefont {Paquette}}, \bibinfo {author} {\bibfnamefont {M.}~\bibnamefont
  {Pelstring}}, \bibinfo {author} {\bibfnamefont {A.}~\bibnamefont
  {Polloreno}}, \bibinfo {author} {\bibfnamefont {V.}~\bibnamefont {Rawat}},
  \bibinfo {author} {\bibfnamefont {C.~A.}\ \bibnamefont {Ryan}}, \bibinfo
  {author} {\bibfnamefont {R.}~\bibnamefont {Renzas}}, \bibinfo {author}
  {\bibfnamefont {N.}~\bibnamefont {Rubin}}, \bibinfo {author} {\bibfnamefont
  {D.}~\bibnamefont {Russel}}, \bibinfo {author} {\bibfnamefont
  {M.}~\bibnamefont {Rust}}, \bibinfo {author} {\bibfnamefont {D.}~\bibnamefont
  {Scarabelli}}, \bibinfo {author} {\bibfnamefont {M.}~\bibnamefont
  {Selvanayagam}}, \bibinfo {author} {\bibfnamefont {R.}~\bibnamefont
  {Sinclair}}, \bibinfo {author} {\bibfnamefont {R.}~\bibnamefont {Smith}},
  \bibinfo {author} {\bibfnamefont {M.}~\bibnamefont {Suska}}, \bibinfo
  {author} {\bibfnamefont {T.-W.}\ \bibnamefont {To}}, \bibinfo {author}
  {\bibfnamefont {M.}~\bibnamefont {Vahidpour}}, \bibinfo {author}
  {\bibfnamefont {N.}~\bibnamefont {Vodrahalli}}, \bibinfo {author}
  {\bibfnamefont {T.}~\bibnamefont {Whyland}}, \bibinfo {author} {\bibfnamefont
  {K.}~\bibnamefont {Yadav}}, \bibinfo {author} {\bibfnamefont
  {W.}~\bibnamefont {Zeng}},\ and\ \bibinfo {author} {\bibfnamefont {C.~T.}\
  \bibnamefont {Rigetti}},\ }\bibfield  {title} {\bibinfo {title}
  {\textit{Demonstration of Universal Parametric Entangling Gates on a
  Multi-Qubit Lattice}},\ }\href {https://doi.org/10.1126/sciadv.aao3603}
  {\bibfield  {journal} {\bibinfo  {journal} {Sci. Adv.}\ }\textbf {\bibinfo
  {volume} {4}},\ \bibinfo {pages} {eaao3603} (\bibinfo {year}
  {2018})}\BibitemShut {NoStop}%
\bibitem [{\citenamefont {Hutchings}\ \emph {et~al.}(2017)\citenamefont
  {Hutchings}, \citenamefont {Hertzberg}, \citenamefont {Liu}, \citenamefont
  {Bronn}, \citenamefont {Keefe}, \citenamefont {Brink}, \citenamefont {Chow},\
  and\ \citenamefont {Plourde}}]{IBM_frequency_tunable_transmon}%
  \BibitemOpen
  \bibfield  {author} {\bibinfo {author} {\bibfnamefont {M.~D.}\ \bibnamefont
  {Hutchings}}, \bibinfo {author} {\bibfnamefont {J.~B.}\ \bibnamefont
  {Hertzberg}}, \bibinfo {author} {\bibfnamefont {Y.}~\bibnamefont {Liu}},
  \bibinfo {author} {\bibfnamefont {N.~T.}\ \bibnamefont {Bronn}}, \bibinfo
  {author} {\bibfnamefont {G.~A.}\ \bibnamefont {Keefe}}, \bibinfo {author}
  {\bibfnamefont {M.}~\bibnamefont {Brink}}, \bibinfo {author} {\bibfnamefont
  {J.~M.}\ \bibnamefont {Chow}},\ and\ \bibinfo {author} {\bibfnamefont
  {B.~L.~T.}\ \bibnamefont {Plourde}},\ }\bibfield  {title} {\bibinfo {title}
  {\textit{Tunable Superconducting Qubits with Flux-Independent Coherence}},\
  }\href {https://doi.org/10.1103/PhysRevApplied.8.044003} {\bibfield
  {journal} {\bibinfo  {journal} {Phys. Rev. Appl.}\ }\textbf {\bibinfo
  {volume} {8}},\ \bibinfo {pages} {044003} (\bibinfo {year}
  {2017})}\BibitemShut {NoStop}%
\bibitem [{\citenamefont {Paladino}\ \emph {et~al.}(2014)\citenamefont
  {Paladino}, \citenamefont {Galperin}, \citenamefont {Falci},\ and\
  \citenamefont {Altshuler}}]{Paladino_TLF_review}%
  \BibitemOpen
  \bibfield  {author} {\bibinfo {author} {\bibfnamefont {E.}~\bibnamefont
  {Paladino}}, \bibinfo {author} {\bibfnamefont {Y.~M.}\ \bibnamefont
  {Galperin}}, \bibinfo {author} {\bibfnamefont {G.}~\bibnamefont {Falci}},\
  and\ \bibinfo {author} {\bibfnamefont {B.~L.}\ \bibnamefont {Altshuler}},\
  }\bibfield  {title} {\bibinfo {title} {$1/f$ \textit{Noise: Implications for
  Solid-State Quantum Information}},\ }\href
  {https://doi.org/10.1103/RevModPhys.86.361} {\bibfield  {journal} {\bibinfo
  {journal} {Rev. Mod. Phys.}\ }\textbf {\bibinfo {volume} {86}},\ \bibinfo
  {pages} {361} (\bibinfo {year} {2014})}\BibitemShut {NoStop}%
\bibitem [{\citenamefont {Kandala}\ \emph {et~al.}(2021)\citenamefont
  {Kandala}, \citenamefont {Wei}, \citenamefont {Srinivasan}, \citenamefont
  {Magesan}, \citenamefont {Carnevale}, \citenamefont {Keefe}, \citenamefont
  {Klaus}, \citenamefont {Dial},\ and\ \citenamefont {McKay}}]{IBM_direct_CR}%
  \BibitemOpen
  \bibfield  {author} {\bibinfo {author} {\bibfnamefont {A.}~\bibnamefont
  {Kandala}}, \bibinfo {author} {\bibfnamefont {K.~X.}\ \bibnamefont {Wei}},
  \bibinfo {author} {\bibfnamefont {S.}~\bibnamefont {Srinivasan}}, \bibinfo
  {author} {\bibfnamefont {E.}~\bibnamefont {Magesan}}, \bibinfo {author}
  {\bibfnamefont {S.}~\bibnamefont {Carnevale}}, \bibinfo {author}
  {\bibfnamefont {G.~A.}\ \bibnamefont {Keefe}}, \bibinfo {author}
  {\bibfnamefont {D.}~\bibnamefont {Klaus}}, \bibinfo {author} {\bibfnamefont
  {O.}~\bibnamefont {Dial}},\ and\ \bibinfo {author} {\bibfnamefont {D.~C.}\
  \bibnamefont {McKay}},\ }\bibfield  {title} {\bibinfo {title}
  {\textit{Demonstration of a High-Fidelity cnot Gate for Fixed-Frequency
  Transmons with Engineered $ZZ$ Suppression}},\ }\href
  {https://doi.org/10.1103/PhysRevLett.127.130501} {\bibfield  {journal}
  {\bibinfo  {journal} {Phys. Rev. Lett.}\ }\textbf {\bibinfo {volume} {127}},\
  \bibinfo {pages} {130501} (\bibinfo {year} {2021})}\BibitemShut {NoStop}%
\bibitem [{\citenamefont {Wei}\ \emph {et~al.}(2022)\citenamefont {Wei},
  \citenamefont {Magesan}, \citenamefont {Lauer}, \citenamefont {Srinivasan},
  \citenamefont {Bogorin}, \citenamefont {Carnevale}, \citenamefont {Keefe},
  \citenamefont {Kim}, \citenamefont {Klaus}, \citenamefont {Landers},
  \citenamefont {Sundaresan}, \citenamefont {Wang}, \citenamefont {Zhang},
  \citenamefont {Steffen}, \citenamefont {Dial}, \citenamefont {McKay},\ and\
  \citenamefont {Kandala}}]{IBM_sizzle}%
  \BibitemOpen
  \bibfield  {author} {\bibinfo {author} {\bibfnamefont {K.~X.}\ \bibnamefont
  {Wei}}, \bibinfo {author} {\bibfnamefont {E.}~\bibnamefont {Magesan}},
  \bibinfo {author} {\bibfnamefont {I.}~\bibnamefont {Lauer}}, \bibinfo
  {author} {\bibfnamefont {S.}~\bibnamefont {Srinivasan}}, \bibinfo {author}
  {\bibfnamefont {D.~F.}\ \bibnamefont {Bogorin}}, \bibinfo {author}
  {\bibfnamefont {S.}~\bibnamefont {Carnevale}}, \bibinfo {author}
  {\bibfnamefont {G.~A.}\ \bibnamefont {Keefe}}, \bibinfo {author}
  {\bibfnamefont {Y.}~\bibnamefont {Kim}}, \bibinfo {author} {\bibfnamefont
  {D.}~\bibnamefont {Klaus}}, \bibinfo {author} {\bibfnamefont
  {W.}~\bibnamefont {Landers}}, \bibinfo {author} {\bibfnamefont
  {N.}~\bibnamefont {Sundaresan}}, \bibinfo {author} {\bibfnamefont
  {C.}~\bibnamefont {Wang}}, \bibinfo {author} {\bibfnamefont {E.~J.}\
  \bibnamefont {Zhang}}, \bibinfo {author} {\bibfnamefont {M.}~\bibnamefont
  {Steffen}}, \bibinfo {author} {\bibfnamefont {O.~E.}\ \bibnamefont {Dial}},
  \bibinfo {author} {\bibfnamefont {D.~C.}\ \bibnamefont {McKay}},\ and\
  \bibinfo {author} {\bibfnamefont {A.}~\bibnamefont {Kandala}},\ }\bibfield
  {title} {\bibinfo {title} {\textit{Hamiltonian Engineering with Multicolor
  Drives for Fast Entangling Gates and Quantum Crosstalk Cancellation}},\
  }\href {https://doi.org/10.1103/PhysRevLett.129.060501} {\bibfield  {journal}
  {\bibinfo  {journal} {Phys. Rev. Lett.}\ }\textbf {\bibinfo {volume} {129}},\
  \bibinfo {pages} {060501} (\bibinfo {year} {2022})}\BibitemShut {NoStop}%
\bibitem [{\citenamefont {Wei}\ \emph {et~al.}(2023)\citenamefont {Wei},
  \citenamefont {Lauer}, \citenamefont {Pritchett}, \citenamefont {Shanks},
  \citenamefont {McKay},\ and\ \citenamefont {Javadi-Abhari}}]{IBM_Bgate}%
  \BibitemOpen
  \bibfield  {author} {\bibinfo {author} {\bibfnamefont {K.~X.}\ \bibnamefont
  {Wei}}, \bibinfo {author} {\bibfnamefont {I.}~\bibnamefont {Lauer}}, \bibinfo
  {author} {\bibfnamefont {E.}~\bibnamefont {Pritchett}}, \bibinfo {author}
  {\bibfnamefont {W.}~\bibnamefont {Shanks}}, \bibinfo {author} {\bibfnamefont
  {D.~C.}\ \bibnamefont {McKay}},\ and\ \bibinfo {author} {\bibfnamefont
  {A.}~\bibnamefont {Javadi-Abhari}},\ }\href@noop {} {\bibinfo {title}
  {\textit{Native Two-Qubit Gates in Fixed-Coupling, Fixed-Frequency Transmons
  Beyond Cross-Resonance Interaction}}} (\bibinfo {year} {2023}),\ \bibinfo
  {note}
  {\href{https://doi.org/10.48550/arXiv.2310.12146}{arXiv:2310.12146}}\BibitemShut
  {NoStop}%
\bibitem [{\citenamefont {Mitchell}\ \emph {et~al.}(2021)\citenamefont
  {Mitchell}, \citenamefont {Naik}, \citenamefont {Morvan}, \citenamefont
  {Hashim}, \citenamefont {Kreikebaum}, \citenamefont {Marinelli},
  \citenamefont {Lavrijsen}, \citenamefont {Nowrouzi}, \citenamefont
  {Santiago},\ and\ \citenamefont {Siddiqi}}]{siddiqi_sizzle}%
  \BibitemOpen
  \bibfield  {author} {\bibinfo {author} {\bibfnamefont {B.~K.}\ \bibnamefont
  {Mitchell}}, \bibinfo {author} {\bibfnamefont {R.~K.}\ \bibnamefont {Naik}},
  \bibinfo {author} {\bibfnamefont {A.}~\bibnamefont {Morvan}}, \bibinfo
  {author} {\bibfnamefont {A.}~\bibnamefont {Hashim}}, \bibinfo {author}
  {\bibfnamefont {J.~M.}\ \bibnamefont {Kreikebaum}}, \bibinfo {author}
  {\bibfnamefont {B.}~\bibnamefont {Marinelli}}, \bibinfo {author}
  {\bibfnamefont {W.}~\bibnamefont {Lavrijsen}}, \bibinfo {author}
  {\bibfnamefont {K.}~\bibnamefont {Nowrouzi}}, \bibinfo {author}
  {\bibfnamefont {D.~I.}\ \bibnamefont {Santiago}},\ and\ \bibinfo {author}
  {\bibfnamefont {I.}~\bibnamefont {Siddiqi}},\ }\bibfield  {title} {\bibinfo
  {title} {\textit{Hardware-Efficient Microwave-Activated Tunable Coupling
  between Superconducting Qubits}},\ }\href
  {https://doi.org/10.1103/PhysRevLett.127.200502} {\bibfield  {journal}
  {\bibinfo  {journal} {Phys. Rev. Lett.}\ }\textbf {\bibinfo {volume} {127}},\
  \bibinfo {pages} {200502} (\bibinfo {year} {2021})}\BibitemShut {NoStop}%
\bibitem [{\citenamefont {Nguyen}\ \emph {et~al.}(2022)\citenamefont {Nguyen},
  \citenamefont {Kim}, \citenamefont {Hashim}, \citenamefont {Goss},
  \citenamefont {Marinelli}, \citenamefont {Bhandari}, \citenamefont {Das},
  \citenamefont {Naik}, \citenamefont {Kreikebaum}, \citenamefont {Jordan},
  \citenamefont {Santiago},\ and\ \citenamefont
  {Siddiqi}}]{Siddiqi_floquet_qubit}%
  \BibitemOpen
  \bibfield  {author} {\bibinfo {author} {\bibfnamefont {L.~B.}\ \bibnamefont
  {Nguyen}}, \bibinfo {author} {\bibfnamefont {Y.}~\bibnamefont {Kim}},
  \bibinfo {author} {\bibfnamefont {A.}~\bibnamefont {Hashim}}, \bibinfo
  {author} {\bibfnamefont {N.}~\bibnamefont {Goss}}, \bibinfo {author}
  {\bibfnamefont {B.}~\bibnamefont {Marinelli}}, \bibinfo {author}
  {\bibfnamefont {B.}~\bibnamefont {Bhandari}}, \bibinfo {author}
  {\bibfnamefont {D.}~\bibnamefont {Das}}, \bibinfo {author} {\bibfnamefont
  {R.~K.}\ \bibnamefont {Naik}}, \bibinfo {author} {\bibfnamefont {J.~M.}\
  \bibnamefont {Kreikebaum}}, \bibinfo {author} {\bibfnamefont {A.~N.}\
  \bibnamefont {Jordan}}, \bibinfo {author} {\bibfnamefont {D.~I.}\
  \bibnamefont {Santiago}},\ and\ \bibinfo {author} {\bibfnamefont
  {I.}~\bibnamefont {Siddiqi}},\ }\href@noop {} {\bibinfo {title}
  {\textit{Programmable Heisenberg Interactions Between Floquet Qubits}}}
  (\bibinfo {year} {2022}),\ \bibinfo {note}
  {\href{https://doi.org/10.48550/arXiv.2211.10383}{arXiv:2211.10383}}\BibitemShut
  {NoStop}%
\bibitem [{\citenamefont {Zhao}(2023)}]{Zhao_bus_fine_tuning}%
  \BibitemOpen
  \bibfield  {author} {\bibinfo {author} {\bibfnamefont {P.}~\bibnamefont
  {Zhao}},\ }\href@noop {} {\bibinfo {title} {\textit{Mitigation of Quantum
  Crosstalk in Cross-Resonance Based Qubit Architectures}}} (\bibinfo {year}
  {2023}),\ \bibinfo {note}
  {\href{https://doi.org/10.48550/arXiv.2307.09995}{arXiv:2307.09995}}\BibitemShut
  {NoStop}%
\bibitem [{\citenamefont {Puri}\ and\ \citenamefont
  {Blais}(2016)}]{Blais_RIP_gate}%
  \BibitemOpen
  \bibfield  {author} {\bibinfo {author} {\bibfnamefont {S.}~\bibnamefont
  {Puri}}\ and\ \bibinfo {author} {\bibfnamefont {A.}~\bibnamefont {Blais}},\
  }\bibfield  {title} {\bibinfo {title} {\textit{High-Fidelity
  Resonator-Induced Phase Gate with Single-Mode Squeezing}},\ }\href
  {https://doi.org/10.1103/PhysRevLett.116.180501} {\bibfield  {journal}
  {\bibinfo  {journal} {Phys. Rev. Lett.}\ }\textbf {\bibinfo {volume} {116}},\
  \bibinfo {pages} {180501} (\bibinfo {year} {2016})}\BibitemShut {NoStop}%
\bibitem [{\citenamefont {Paik}\ \emph {et~al.}(2016)\citenamefont {Paik},
  \citenamefont {Mezzacapo}, \citenamefont {Sandberg}, \citenamefont {McClure},
  \citenamefont {Abdo}, \citenamefont {C\'orcoles}, \citenamefont {Dial},
  \citenamefont {Bogorin}, \citenamefont {Plourde}, \citenamefont {Steffen},
  \citenamefont {Cross}, \citenamefont {Gambetta},\ and\ \citenamefont
  {Chow}}]{IBM_RIP}%
  \BibitemOpen
  \bibfield  {author} {\bibinfo {author} {\bibfnamefont {H.}~\bibnamefont
  {Paik}}, \bibinfo {author} {\bibfnamefont {A.}~\bibnamefont {Mezzacapo}},
  \bibinfo {author} {\bibfnamefont {M.}~\bibnamefont {Sandberg}}, \bibinfo
  {author} {\bibfnamefont {D.~T.}\ \bibnamefont {McClure}}, \bibinfo {author}
  {\bibfnamefont {B.}~\bibnamefont {Abdo}}, \bibinfo {author} {\bibfnamefont
  {A.~D.}\ \bibnamefont {C\'orcoles}}, \bibinfo {author} {\bibfnamefont
  {O.}~\bibnamefont {Dial}}, \bibinfo {author} {\bibfnamefont {D.~F.}\
  \bibnamefont {Bogorin}}, \bibinfo {author} {\bibfnamefont {B.~L.~T.}\
  \bibnamefont {Plourde}}, \bibinfo {author} {\bibfnamefont {M.}~\bibnamefont
  {Steffen}}, \bibinfo {author} {\bibfnamefont {A.~W.}\ \bibnamefont {Cross}},
  \bibinfo {author} {\bibfnamefont {J.~M.}\ \bibnamefont {Gambetta}},\ and\
  \bibinfo {author} {\bibfnamefont {J.~M.}\ \bibnamefont {Chow}},\ }\bibfield
  {title} {\bibinfo {title} {\textit{Experimental Demonstration of a
  Resonator-Induced Phase Gate in a Multiqubit Circuit-QED System}},\ }\href
  {https://doi.org/10.1103/PhysRevLett.117.250502} {\bibfield  {journal}
  {\bibinfo  {journal} {Phys. Rev. Lett.}\ }\textbf {\bibinfo {volume} {117}},\
  \bibinfo {pages} {250502} (\bibinfo {year} {2016})}\BibitemShut {NoStop}%
\bibitem [{\citenamefont {Malekakhlagh}\ \emph {et~al.}(2022)\citenamefont
  {Malekakhlagh}, \citenamefont {Shanks},\ and\ \citenamefont
  {Paik}}]{IBM_RIP_leakage}%
  \BibitemOpen
  \bibfield  {author} {\bibinfo {author} {\bibfnamefont {M.}~\bibnamefont
  {Malekakhlagh}}, \bibinfo {author} {\bibfnamefont {W.}~\bibnamefont
  {Shanks}},\ and\ \bibinfo {author} {\bibfnamefont {H.}~\bibnamefont {Paik}},\
  }\bibfield  {title} {\bibinfo {title} {\textit{Optimization of the
  Resonator-Induced Phase Gate for Superconducting Qubits}},\ }\href
  {https://doi.org/10.1103/PhysRevA.105.022607} {\bibfield  {journal} {\bibinfo
   {journal} {Phys. Rev. A}\ }\textbf {\bibinfo {volume} {105}},\ \bibinfo
  {pages} {022607} (\bibinfo {year} {2022})}\BibitemShut {NoStop}%
\bibitem [{\citenamefont {Cross}\ and\ \citenamefont
  {Gambetta}(2015)}]{IBM_RIP_optimization}%
  \BibitemOpen
  \bibfield  {author} {\bibinfo {author} {\bibfnamefont {A.~W.}\ \bibnamefont
  {Cross}}\ and\ \bibinfo {author} {\bibfnamefont {J.~M.}\ \bibnamefont
  {Gambetta}},\ }\bibfield  {title} {\bibinfo {title} {\textit{Optimized Pulse
  Shapes for a Resonator-Induced Phase Gate}},\ }\href
  {https://doi.org/10.1103/PhysRevA.91.032325} {\bibfield  {journal} {\bibinfo
  {journal} {Phys. Rev. A}\ }\textbf {\bibinfo {volume} {91}},\ \bibinfo
  {pages} {032325} (\bibinfo {year} {2015})}\BibitemShut {NoStop}%
\bibitem [{\citenamefont {Romanenko}\ \emph {et~al.}(2020)\citenamefont
  {Romanenko}, \citenamefont {Pilipenko}, \citenamefont {Zorzetti},
  \citenamefont {Frolov}, \citenamefont {Awida}, \citenamefont {Belomestnykh},
  \citenamefont {Posen},\ and\ \citenamefont
  {Grassellino}}]{Grassellino_Cavity}%
  \BibitemOpen
  \bibfield  {author} {\bibinfo {author} {\bibfnamefont {A.}~\bibnamefont
  {Romanenko}}, \bibinfo {author} {\bibfnamefont {R.}~\bibnamefont
  {Pilipenko}}, \bibinfo {author} {\bibfnamefont {S.}~\bibnamefont {Zorzetti}},
  \bibinfo {author} {\bibfnamefont {D.}~\bibnamefont {Frolov}}, \bibinfo
  {author} {\bibfnamefont {M.}~\bibnamefont {Awida}}, \bibinfo {author}
  {\bibfnamefont {S.}~\bibnamefont {Belomestnykh}}, \bibinfo {author}
  {\bibfnamefont {S.}~\bibnamefont {Posen}},\ and\ \bibinfo {author}
  {\bibfnamefont {A.}~\bibnamefont {Grassellino}},\ }\bibfield  {title}
  {\bibinfo {title} {\textit{Three-Dimensional Superconducting Resonators at
  $20$ mK with Photon Lifetimes up to $\ensuremath{\tau}=2$ s}},\ }\href
  {https://doi.org/10.1103/PhysRevApplied.13.034032} {\bibfield  {journal}
  {\bibinfo  {journal} {Phys. Rev. Appl.}\ }\textbf {\bibinfo {volume} {13}},\
  \bibinfo {pages} {034032} (\bibinfo {year} {2020})}\BibitemShut {NoStop}%
\bibitem [{\citenamefont {Milul}\ \emph {et~al.}(2023)\citenamefont {Milul},
  \citenamefont {Guttel}, \citenamefont {Goldblatt}, \citenamefont {Hazanov},
  \citenamefont {Joshi}, \citenamefont {Chausovsky}, \citenamefont {Kahn},
  \citenamefont {\ifmmode~\mbox{\c{C}}\else \c{C}\fi{}ifty\"urek},
  \citenamefont {Lafont},\ and\ \citenamefont {Rosenblum}}]{Rosenblum_cavity}%
  \BibitemOpen
  \bibfield  {author} {\bibinfo {author} {\bibfnamefont {O.}~\bibnamefont
  {Milul}}, \bibinfo {author} {\bibfnamefont {B.}~\bibnamefont {Guttel}},
  \bibinfo {author} {\bibfnamefont {U.}~\bibnamefont {Goldblatt}}, \bibinfo
  {author} {\bibfnamefont {S.}~\bibnamefont {Hazanov}}, \bibinfo {author}
  {\bibfnamefont {L.~M.}\ \bibnamefont {Joshi}}, \bibinfo {author}
  {\bibfnamefont {D.}~\bibnamefont {Chausovsky}}, \bibinfo {author}
  {\bibfnamefont {N.}~\bibnamefont {Kahn}}, \bibinfo {author} {\bibfnamefont
  {E.}~\bibnamefont {\ifmmode~\mbox{\c{C}}\else \c{C}\fi{}ifty\"urek}},
  \bibinfo {author} {\bibfnamefont {F.}~\bibnamefont {Lafont}},\ and\ \bibinfo
  {author} {\bibfnamefont {S.}~\bibnamefont {Rosenblum}},\ }\bibfield  {title}
  {\bibinfo {title} {\textit{Superconducting Cavity Qubit with Tens of
  Milliseconds Single-Photon Coherence Time}},\ }\href
  {https://doi.org/10.1103/PRXQuantum.4.030336} {\bibfield  {journal} {\bibinfo
   {journal} {PRX Quantum}\ }\textbf {\bibinfo {volume} {4}},\ \bibinfo {pages}
  {030336} (\bibinfo {year} {2023})}\BibitemShut {NoStop}%
\bibitem [{\citenamefont {Ganjam}\ \emph {et~al.}(2023)\citenamefont {Ganjam},
  \citenamefont {Wang}, \citenamefont {Lu}, \citenamefont {Banerjee},
  \citenamefont {Lei}, \citenamefont {Krayzman}, \citenamefont {Kisslinger},
  \citenamefont {Zhou}, \citenamefont {Li}, \citenamefont {Jia}, \citenamefont
  {Liu}, \citenamefont {Frunzio},\ and\ \citenamefont
  {Schoelkopf}}]{Yale_2d_resonator}%
  \BibitemOpen
  \bibfield  {author} {\bibinfo {author} {\bibfnamefont {S.}~\bibnamefont
  {Ganjam}}, \bibinfo {author} {\bibfnamefont {Y.}~\bibnamefont {Wang}},
  \bibinfo {author} {\bibfnamefont {Y.}~\bibnamefont {Lu}}, \bibinfo {author}
  {\bibfnamefont {A.}~\bibnamefont {Banerjee}}, \bibinfo {author}
  {\bibfnamefont {C.~U.}\ \bibnamefont {Lei}}, \bibinfo {author} {\bibfnamefont
  {L.}~\bibnamefont {Krayzman}}, \bibinfo {author} {\bibfnamefont
  {K.}~\bibnamefont {Kisslinger}}, \bibinfo {author} {\bibfnamefont
  {C.}~\bibnamefont {Zhou}}, \bibinfo {author} {\bibfnamefont {R.}~\bibnamefont
  {Li}}, \bibinfo {author} {\bibfnamefont {Y.}~\bibnamefont {Jia}}, \bibinfo
  {author} {\bibfnamefont {M.}~\bibnamefont {Liu}}, \bibinfo {author}
  {\bibfnamefont {L.}~\bibnamefont {Frunzio}},\ and\ \bibinfo {author}
  {\bibfnamefont {R.~J.}\ \bibnamefont {Schoelkopf}},\ }\href@noop {} {\bibinfo
  {title} {\textit{Surpassing Millisecond Coherence Times in On-Chip
  Superconducting Quantum Memories by Optimizing Materials, Processes, and
  Circuit Design}}} (\bibinfo {year} {2023}),\ \bibinfo {note}
  {\href{https://doi.org/10.48550/arXiv.2308.15539}{arXiv:2308.15539}}\BibitemShut
  {NoStop}%
\bibitem [{\citenamefont {Paraoanu}(2006)}]{Paraoanu_CR}%
  \BibitemOpen
  \bibfield  {author} {\bibinfo {author} {\bibfnamefont {G.~S.}\ \bibnamefont
  {Paraoanu}},\ }\bibfield  {title} {\bibinfo {title}
  {\textit{Microwave-Induced Coupling of Superconducting Qubits}},\ }\href
  {https://doi.org/10.1103/PhysRevB.74.140504} {\bibfield  {journal} {\bibinfo
  {journal} {Phys. Rev. B}\ }\textbf {\bibinfo {volume} {74}},\ \bibinfo
  {pages} {140504} (\bibinfo {year} {2006})}\BibitemShut {NoStop}%
\bibitem [{\citenamefont {Sheldon}\ \emph {et~al.}(2016)\citenamefont
  {Sheldon}, \citenamefont {Magesan}, \citenamefont {Chow},\ and\ \citenamefont
  {Gambetta}}]{Sheldon_CR}%
  \BibitemOpen
  \bibfield  {author} {\bibinfo {author} {\bibfnamefont {S.}~\bibnamefont
  {Sheldon}}, \bibinfo {author} {\bibfnamefont {E.}~\bibnamefont {Magesan}},
  \bibinfo {author} {\bibfnamefont {J.~M.}\ \bibnamefont {Chow}},\ and\
  \bibinfo {author} {\bibfnamefont {J.~M.}\ \bibnamefont {Gambetta}},\
  }\bibfield  {title} {\bibinfo {title} {\textit{Procedure for Systematically
  Tuning up Cross-Talk in the Cross-Resonance Gate}},\ }\href
  {https://doi.org/10.1103/PhysRevA.93.060302} {\bibfield  {journal} {\bibinfo
  {journal} {Phys. Rev. A}\ }\textbf {\bibinfo {volume} {93}},\ \bibinfo
  {pages} {060302} (\bibinfo {year} {2016})}\BibitemShut {NoStop}%
\bibitem [{\citenamefont {Rigetti}\ and\ \citenamefont
  {Devoret}(2010{\natexlab{a}})}]{Devoret_Rigetti_CR}%
  \BibitemOpen
  \bibfield  {author} {\bibinfo {author} {\bibfnamefont {C.}~\bibnamefont
  {Rigetti}}\ and\ \bibinfo {author} {\bibfnamefont {M.}~\bibnamefont
  {Devoret}},\ }\bibfield  {title} {\bibinfo {title} {\textit{Fully
  Microwave-Tunable Universal Gates in Superconducting Qubits with Linear
  Couplings and Fixed Transition Frequencies}},\ }\href
  {https://doi.org/10.1103/PhysRevB.81.134507} {\bibfield  {journal} {\bibinfo
  {journal} {Phys. Rev. B}\ }\textbf {\bibinfo {volume} {81}},\ \bibinfo
  {pages} {134507} (\bibinfo {year} {2010}{\natexlab{a}})}\BibitemShut
  {NoStop}%
\bibitem [{\citenamefont {Sundaresan}\ \emph {et~al.}(2020)\citenamefont
  {Sundaresan}, \citenamefont {Lauer}, \citenamefont {Pritchett}, \citenamefont
  {Magesan}, \citenamefont {Jurcevic},\ and\ \citenamefont
  {Gambetta}}]{IBM_echo_CR}%
  \BibitemOpen
  \bibfield  {author} {\bibinfo {author} {\bibfnamefont {N.}~\bibnamefont
  {Sundaresan}}, \bibinfo {author} {\bibfnamefont {I.}~\bibnamefont {Lauer}},
  \bibinfo {author} {\bibfnamefont {E.}~\bibnamefont {Pritchett}}, \bibinfo
  {author} {\bibfnamefont {E.}~\bibnamefont {Magesan}}, \bibinfo {author}
  {\bibfnamefont {P.}~\bibnamefont {Jurcevic}},\ and\ \bibinfo {author}
  {\bibfnamefont {J.~M.}\ \bibnamefont {Gambetta}},\ }\bibfield  {title}
  {\bibinfo {title} {\textit{Reducing Unitary and Spectator Errors in Cross
  Resonance with Optimized Rotary Echoes}},\ }\href
  {https://doi.org/10.1103/PRXQuantum.1.020318} {\bibfield  {journal} {\bibinfo
   {journal} {PRX Quantum}\ }\textbf {\bibinfo {volume} {1}},\ \bibinfo {pages}
  {020318} (\bibinfo {year} {2020})}\BibitemShut {NoStop}%
\bibitem [{\citenamefont {Tripathi}\ \emph {et~al.}(2019)\citenamefont
  {Tripathi}, \citenamefont {Khezri},\ and\ \citenamefont
  {Korotkov}}]{Riverside_CR_errorbudget}%
  \BibitemOpen
  \bibfield  {author} {\bibinfo {author} {\bibfnamefont {V.}~\bibnamefont
  {Tripathi}}, \bibinfo {author} {\bibfnamefont {M.}~\bibnamefont {Khezri}},\
  and\ \bibinfo {author} {\bibfnamefont {A.~N.}\ \bibnamefont {Korotkov}},\
  }\bibfield  {title} {\bibinfo {title} {\textit{Operation and Intrinsic Error
  Budget of a Two-Qubit Cross-Resonance Gate}},\ }\href
  {https://doi.org/10.1103/PhysRevA.100.012301} {\bibfield  {journal} {\bibinfo
   {journal} {Phys. Rev. A}\ }\textbf {\bibinfo {volume} {100}},\ \bibinfo
  {pages} {012301} (\bibinfo {year} {2019})}\BibitemShut {NoStop}%
\bibitem [{\citenamefont {Magesan}\ and\ \citenamefont
  {Gambetta}(2020)}]{IBM_CR_effective}%
  \BibitemOpen
  \bibfield  {author} {\bibinfo {author} {\bibfnamefont {E.}~\bibnamefont
  {Magesan}}\ and\ \bibinfo {author} {\bibfnamefont {J.~M.}\ \bibnamefont
  {Gambetta}},\ }\bibfield  {title} {\bibinfo {title} {\textit{Effective
  Hamiltonian Models of the Cross-Resonance Gate}},\ }\href
  {https://doi.org/10.1103/PhysRevA.101.052308} {\bibfield  {journal} {\bibinfo
   {journal} {Phys. Rev. A}\ }\textbf {\bibinfo {volume} {101}},\ \bibinfo
  {pages} {052308} (\bibinfo {year} {2020})}\BibitemShut {NoStop}%
\bibitem [{\citenamefont {Paik}\ \emph {et~al.}(2011)\citenamefont {Paik},
  \citenamefont {Schuster}, \citenamefont {Bishop}, \citenamefont {Kirchmair},
  \citenamefont {Catelani}, \citenamefont {Sears}, \citenamefont {Johnson},
  \citenamefont {Reagor}, \citenamefont {Frunzio}, \citenamefont {Glazman},
  \citenamefont {Girvin}, \citenamefont {Devoret},\ and\ \citenamefont
  {Schoelkopf}}]{Schoelkopf_3d_transmon}%
  \BibitemOpen
  \bibfield  {author} {\bibinfo {author} {\bibfnamefont {H.}~\bibnamefont
  {Paik}}, \bibinfo {author} {\bibfnamefont {D.~I.}\ \bibnamefont {Schuster}},
  \bibinfo {author} {\bibfnamefont {L.~S.}\ \bibnamefont {Bishop}}, \bibinfo
  {author} {\bibfnamefont {G.}~\bibnamefont {Kirchmair}}, \bibinfo {author}
  {\bibfnamefont {G.}~\bibnamefont {Catelani}}, \bibinfo {author}
  {\bibfnamefont {A.~P.}\ \bibnamefont {Sears}}, \bibinfo {author}
  {\bibfnamefont {B.~R.}\ \bibnamefont {Johnson}}, \bibinfo {author}
  {\bibfnamefont {M.~J.}\ \bibnamefont {Reagor}}, \bibinfo {author}
  {\bibfnamefont {L.}~\bibnamefont {Frunzio}}, \bibinfo {author} {\bibfnamefont
  {L.~I.}\ \bibnamefont {Glazman}}, \bibinfo {author} {\bibfnamefont {S.~M.}\
  \bibnamefont {Girvin}}, \bibinfo {author} {\bibfnamefont {M.~H.}\
  \bibnamefont {Devoret}},\ and\ \bibinfo {author} {\bibfnamefont {R.~J.}\
  \bibnamefont {Schoelkopf}},\ }\bibfield  {title} {\bibinfo {title}
  {\textit{Observation of High Coherence in Josephson Junction Qubits Measured
  in a Three-Dimensional Circuit QED Architecture}},\ }\href
  {https://doi.org/10.1103/PhysRevLett.107.240501} {\bibfield  {journal}
  {\bibinfo  {journal} {Phys. Rev. Lett.}\ }\textbf {\bibinfo {volume} {107}},\
  \bibinfo {pages} {240501} (\bibinfo {year} {2011})}\BibitemShut {NoStop}%
\bibitem [{\citenamefont {Place}\ \emph {et~al.}(2021)\citenamefont {Place},
  \citenamefont {Rodgers}, \citenamefont {Mundada}, \citenamefont {Smitham},
  \citenamefont {Fitzpatrick}, \citenamefont {Leng}, \citenamefont {Premkumar},
  \citenamefont {Bryon}, \citenamefont {Vrajitoarea}, \citenamefont {Sussman},
  \citenamefont {Cheng}, \citenamefont {Madhavan}, \citenamefont {Babla},
  \citenamefont {Le}, \citenamefont {Gang}, \citenamefont {J{\"a}ck},
  \citenamefont {Gyenis}, \citenamefont {Yao}, \citenamefont {Cava},
  \citenamefont {de~Leon},\ and\ \citenamefont {Houck}}]{Houck_tantalum_qubit}%
  \BibitemOpen
  \bibfield  {author} {\bibinfo {author} {\bibfnamefont {A.~P.~M.}\
  \bibnamefont {Place}}, \bibinfo {author} {\bibfnamefont {L.~V.~H.}\
  \bibnamefont {Rodgers}}, \bibinfo {author} {\bibfnamefont {P.}~\bibnamefont
  {Mundada}}, \bibinfo {author} {\bibfnamefont {B.~M.}\ \bibnamefont
  {Smitham}}, \bibinfo {author} {\bibfnamefont {M.}~\bibnamefont
  {Fitzpatrick}}, \bibinfo {author} {\bibfnamefont {Z.}~\bibnamefont {Leng}},
  \bibinfo {author} {\bibfnamefont {A.}~\bibnamefont {Premkumar}}, \bibinfo
  {author} {\bibfnamefont {J.}~\bibnamefont {Bryon}}, \bibinfo {author}
  {\bibfnamefont {A.}~\bibnamefont {Vrajitoarea}}, \bibinfo {author}
  {\bibfnamefont {S.}~\bibnamefont {Sussman}}, \bibinfo {author} {\bibfnamefont
  {G.}~\bibnamefont {Cheng}}, \bibinfo {author} {\bibfnamefont
  {T.}~\bibnamefont {Madhavan}}, \bibinfo {author} {\bibfnamefont {H.~K.}\
  \bibnamefont {Babla}}, \bibinfo {author} {\bibfnamefont {X.~H.}\ \bibnamefont
  {Le}}, \bibinfo {author} {\bibfnamefont {Y.}~\bibnamefont {Gang}}, \bibinfo
  {author} {\bibfnamefont {B.}~\bibnamefont {J{\"a}ck}}, \bibinfo {author}
  {\bibfnamefont {A.}~\bibnamefont {Gyenis}}, \bibinfo {author} {\bibfnamefont
  {N.}~\bibnamefont {Yao}}, \bibinfo {author} {\bibfnamefont {R.~J.}\
  \bibnamefont {Cava}}, \bibinfo {author} {\bibfnamefont {N.~P.}\ \bibnamefont
  {de~Leon}},\ and\ \bibinfo {author} {\bibfnamefont {A.~A.}\ \bibnamefont
  {Houck}},\ }\bibfield  {title} {\bibinfo {title} {\textit{New Material
  Platform for Superconducting Transmon Qubits with Coherence Times Exceeding
  0.3 Milliseconds}},\ }\href {https://doi.org/10.1038/s41467-021-22030-5}
  {\bibfield  {journal} {\bibinfo  {journal} {Nat. Commun.}\ }\textbf {\bibinfo
  {volume} {12}},\ \bibinfo {pages} {1779} (\bibinfo {year}
  {2021})}\BibitemShut {NoStop}%
\bibitem [{\citenamefont {Wang}\ \emph {et~al.}(2022)\citenamefont {Wang},
  \citenamefont {Li}, \citenamefont {Xu}, \citenamefont {Li}, \citenamefont
  {Wang}, \citenamefont {Yang}, \citenamefont {Mi}, \citenamefont {Liang},
  \citenamefont {Su}, \citenamefont {Yang}, \citenamefont {Wang}, \citenamefont
  {Wang}, \citenamefont {Li}, \citenamefont {Chen}, \citenamefont {Li},
  \citenamefont {Linghu}, \citenamefont {Han}, \citenamefont {Zhang},
  \citenamefont {Feng}, \citenamefont {Song}, \citenamefont {Ma}, \citenamefont
  {Zhang}, \citenamefont {Wang}, \citenamefont {Zhao}, \citenamefont {Liu},
  \citenamefont {Xue}, \citenamefont {Jin},\ and\ \citenamefont
  {Yu}}]{Yu_tantalum_qubit}%
  \BibitemOpen
  \bibfield  {author} {\bibinfo {author} {\bibfnamefont {C.}~\bibnamefont
  {Wang}}, \bibinfo {author} {\bibfnamefont {X.}~\bibnamefont {Li}}, \bibinfo
  {author} {\bibfnamefont {H.}~\bibnamefont {Xu}}, \bibinfo {author}
  {\bibfnamefont {Z.}~\bibnamefont {Li}}, \bibinfo {author} {\bibfnamefont
  {J.}~\bibnamefont {Wang}}, \bibinfo {author} {\bibfnamefont {Z.}~\bibnamefont
  {Yang}}, \bibinfo {author} {\bibfnamefont {Z.}~\bibnamefont {Mi}}, \bibinfo
  {author} {\bibfnamefont {X.}~\bibnamefont {Liang}}, \bibinfo {author}
  {\bibfnamefont {T.}~\bibnamefont {Su}}, \bibinfo {author} {\bibfnamefont
  {C.}~\bibnamefont {Yang}}, \bibinfo {author} {\bibfnamefont {G.}~\bibnamefont
  {Wang}}, \bibinfo {author} {\bibfnamefont {W.}~\bibnamefont {Wang}}, \bibinfo
  {author} {\bibfnamefont {Y.}~\bibnamefont {Li}}, \bibinfo {author}
  {\bibfnamefont {M.}~\bibnamefont {Chen}}, \bibinfo {author} {\bibfnamefont
  {C.}~\bibnamefont {Li}}, \bibinfo {author} {\bibfnamefont {K.}~\bibnamefont
  {Linghu}}, \bibinfo {author} {\bibfnamefont {J.}~\bibnamefont {Han}},
  \bibinfo {author} {\bibfnamefont {Y.}~\bibnamefont {Zhang}}, \bibinfo
  {author} {\bibfnamefont {Y.}~\bibnamefont {Feng}}, \bibinfo {author}
  {\bibfnamefont {Y.}~\bibnamefont {Song}}, \bibinfo {author} {\bibfnamefont
  {T.}~\bibnamefont {Ma}}, \bibinfo {author} {\bibfnamefont {J.}~\bibnamefont
  {Zhang}}, \bibinfo {author} {\bibfnamefont {R.}~\bibnamefont {Wang}},
  \bibinfo {author} {\bibfnamefont {P.}~\bibnamefont {Zhao}}, \bibinfo {author}
  {\bibfnamefont {W.}~\bibnamefont {Liu}}, \bibinfo {author} {\bibfnamefont
  {G.}~\bibnamefont {Xue}}, \bibinfo {author} {\bibfnamefont {Y.}~\bibnamefont
  {Jin}},\ and\ \bibinfo {author} {\bibfnamefont {H.}~\bibnamefont {Yu}},\
  }\bibfield  {title} {\bibinfo {title} {\textit{Towards Practical Quantum
  Computers: Transmon Qubit with a Lifetime Approaching 0.5 Milliseconds}},\
  }\href {https://doi.org/10.1038/s41534-021-00510-2} {\bibfield  {journal}
  {\bibinfo  {journal} {npj Quantum Inf.}\ }\textbf {\bibinfo {volume} {8}},\
  \bibinfo {pages} {3} (\bibinfo {year} {2022})}\BibitemShut {NoStop}%
\bibitem [{\citenamefont {Bal}\ \emph {et~al.}(2023)\citenamefont {Bal},
  \citenamefont {Murthy}, \citenamefont {Zhu}, \citenamefont {Crisa},
  \citenamefont {You}, \citenamefont {Huang}, \citenamefont {Tanay~Roy},
  \citenamefont {van Zanten}, \citenamefont {Pilipenko}, \citenamefont
  {Nekrashevich}, \citenamefont {Bafia}, \citenamefont {Krasnikova},
  \citenamefont {Kopas}, \citenamefont {Lachman}, \citenamefont {Miller},
  \citenamefont {Mutus}, \citenamefont {Reagor}, \citenamefont {Cansizoglu},
  \citenamefont {Marshall}, \citenamefont {Pappas}, \citenamefont {Vu},
  \citenamefont {Yadavalli}, \citenamefont {Oh}, \citenamefont {Zhou},
  \citenamefont {Kramer}, \citenamefont {Lecocq}, \citenamefont {Goronzy},
  \citenamefont {Torres-Castanedo}, \citenamefont {Pritchard}, \citenamefont
  {Dravid}, \citenamefont {Rondinelli}, \citenamefont {Bedzyk}, \citenamefont
  {Hersam}, \citenamefont {Zasadzinski}, \citenamefont {Koch}, \citenamefont
  {Sauls}, \citenamefont {Romanenko},\ and\ \citenamefont
  {Grassellino}}]{Fermilab_encapsulation}%
  \BibitemOpen
  \bibfield  {author} {\bibinfo {author} {\bibfnamefont {M.}~\bibnamefont
  {Bal}}, \bibinfo {author} {\bibfnamefont {A.~A.}\ \bibnamefont {Murthy}},
  \bibinfo {author} {\bibfnamefont {S.}~\bibnamefont {Zhu}}, \bibinfo {author}
  {\bibfnamefont {F.}~\bibnamefont {Crisa}}, \bibinfo {author} {\bibfnamefont
  {X.}~\bibnamefont {You}}, \bibinfo {author} {\bibfnamefont {Z.}~\bibnamefont
  {Huang}}, \bibinfo {author} {\bibfnamefont {J.~L.}\ \bibnamefont
  {Tanay~Roy}}, \bibinfo {author} {\bibfnamefont {D.}~\bibnamefont {van
  Zanten}}, \bibinfo {author} {\bibfnamefont {R.}~\bibnamefont {Pilipenko}},
  \bibinfo {author} {\bibfnamefont {I.}~\bibnamefont {Nekrashevich}}, \bibinfo
  {author} {\bibfnamefont {D.}~\bibnamefont {Bafia}}, \bibinfo {author}
  {\bibfnamefont {Y.}~\bibnamefont {Krasnikova}}, \bibinfo {author}
  {\bibfnamefont {C.~J.}\ \bibnamefont {Kopas}}, \bibinfo {author}
  {\bibfnamefont {E.~O.}\ \bibnamefont {Lachman}}, \bibinfo {author}
  {\bibfnamefont {D.}~\bibnamefont {Miller}}, \bibinfo {author} {\bibfnamefont
  {J.~Y.}\ \bibnamefont {Mutus}}, \bibinfo {author} {\bibfnamefont {M.~J.}\
  \bibnamefont {Reagor}}, \bibinfo {author} {\bibfnamefont {H.}~\bibnamefont
  {Cansizoglu}}, \bibinfo {author} {\bibfnamefont {J.}~\bibnamefont
  {Marshall}}, \bibinfo {author} {\bibfnamefont {D.~P.}\ \bibnamefont
  {Pappas}}, \bibinfo {author} {\bibfnamefont {K.}~\bibnamefont {Vu}}, \bibinfo
  {author} {\bibfnamefont {K.}~\bibnamefont {Yadavalli}}, \bibinfo {author}
  {\bibfnamefont {J.-S.}\ \bibnamefont {Oh}}, \bibinfo {author} {\bibfnamefont
  {L.}~\bibnamefont {Zhou}}, \bibinfo {author} {\bibfnamefont {M.~J.}\
  \bibnamefont {Kramer}}, \bibinfo {author} {\bibfnamefont {F.~Q.}\
  \bibnamefont {Lecocq}}, \bibinfo {author} {\bibfnamefont {D.~P.}\
  \bibnamefont {Goronzy}}, \bibinfo {author} {\bibfnamefont {C.~G.}\
  \bibnamefont {Torres-Castanedo}}, \bibinfo {author} {\bibfnamefont
  {G.}~\bibnamefont {Pritchard}}, \bibinfo {author} {\bibfnamefont {V.~P.}\
  \bibnamefont {Dravid}}, \bibinfo {author} {\bibfnamefont {J.~M.}\
  \bibnamefont {Rondinelli}}, \bibinfo {author} {\bibfnamefont {M.~J.}\
  \bibnamefont {Bedzyk}}, \bibinfo {author} {\bibfnamefont {M.~C.}\
  \bibnamefont {Hersam}}, \bibinfo {author} {\bibfnamefont {J.}~\bibnamefont
  {Zasadzinski}}, \bibinfo {author} {\bibfnamefont {J.}~\bibnamefont {Koch}},
  \bibinfo {author} {\bibfnamefont {J.~A.}\ \bibnamefont {Sauls}}, \bibinfo
  {author} {\bibfnamefont {A.}~\bibnamefont {Romanenko}},\ and\ \bibinfo
  {author} {\bibfnamefont {A.}~\bibnamefont {Grassellino}},\ }\href@noop {}
  {\bibinfo {title} {\textit{Systematic Improvements in Transmon Qubit
  Coherence Enabled by Niobium Surface Encapsulation}}} (\bibinfo {year}
  {2023}),\ \bibinfo {note}
  {\href{https://doi.org/10.48550/arXiv.2304.13257}{arXiv:2304.13257}}\BibitemShut
  {NoStop}%
\bibitem [{\citenamefont {Bizn\'arov\'a}\ \emph {et~al.}(2023)\citenamefont
  {Bizn\'arov\'a}, \citenamefont {Osman}, \citenamefont {Rehnman},
  \citenamefont {Chayanun}, \citenamefont {Kri\v{z}an}, \citenamefont
  {Malmberg}, \citenamefont {Rommel}, \citenamefont {Warren}, \citenamefont
  {Delsing}, \citenamefont {Yurgens}, \citenamefont {Bylander},\ and\
  \citenamefont {Roudsari}}]{Roudsari_transmon_interfacial}%
  \BibitemOpen
  \bibfield  {author} {\bibinfo {author} {\bibfnamefont {J.}~\bibnamefont
  {Bizn\'arov\'a}}, \bibinfo {author} {\bibfnamefont {A.}~\bibnamefont
  {Osman}}, \bibinfo {author} {\bibfnamefont {E.}~\bibnamefont {Rehnman}},
  \bibinfo {author} {\bibfnamefont {L.}~\bibnamefont {Chayanun}}, \bibinfo
  {author} {\bibfnamefont {C.}~\bibnamefont {Kri\v{z}an}}, \bibinfo {author}
  {\bibfnamefont {P.}~\bibnamefont {Malmberg}}, \bibinfo {author}
  {\bibfnamefont {M.}~\bibnamefont {Rommel}}, \bibinfo {author} {\bibfnamefont
  {C.}~\bibnamefont {Warren}}, \bibinfo {author} {\bibfnamefont
  {P.}~\bibnamefont {Delsing}}, \bibinfo {author} {\bibfnamefont
  {A.}~\bibnamefont {Yurgens}}, \bibinfo {author} {\bibfnamefont
  {J.}~\bibnamefont {Bylander}},\ and\ \bibinfo {author} {\bibfnamefont
  {A.~F.}\ \bibnamefont {Roudsari}},\ }\href@noop {} {\bibinfo {title}
  {\textit{Mitigation of Interfacial Dielectric Loss in Aluminum-on-Silicon
  Superconducting Qubits}}} (\bibinfo {year} {2023}),\ \bibinfo {note}
  {\href{https://doi.org/10.48550/arXiv.2310.06797}{arXiv.2310.06797}}\BibitemShut
  {NoStop}%
\bibitem [{\citenamefont {Carroll}\ \emph {et~al.}(2022)\citenamefont
  {Carroll}, \citenamefont {Rosenblatt}, \citenamefont {Jurcevic},
  \citenamefont {Lauer},\ and\ \citenamefont {Kandala}}]{ibm_t1_fluc}%
  \BibitemOpen
  \bibfield  {author} {\bibinfo {author} {\bibfnamefont {M.}~\bibnamefont
  {Carroll}}, \bibinfo {author} {\bibfnamefont {S.}~\bibnamefont {Rosenblatt}},
  \bibinfo {author} {\bibfnamefont {P.}~\bibnamefont {Jurcevic}}, \bibinfo
  {author} {\bibfnamefont {I.}~\bibnamefont {Lauer}},\ and\ \bibinfo {author}
  {\bibfnamefont {A.}~\bibnamefont {Kandala}},\ }\bibfield  {title} {\bibinfo
  {title} {\textit{Dynamics of Superconducting Qubit Relaxation Times}},\
  }\href {https://doi.org/10.1038/s41534-022-00643-y} {\bibfield  {journal}
  {\bibinfo  {journal} {npj Quantum Inf.}\ }\textbf {\bibinfo {volume} {8}},\
  \bibinfo {pages} {132} (\bibinfo {year} {2022})}\BibitemShut {NoStop}%
\bibitem [{\citenamefont {McKay}\ \emph {et~al.}(2016)\citenamefont {McKay},
  \citenamefont {Filipp}, \citenamefont {Mezzacapo}, \citenamefont {Magesan},
  \citenamefont {Chow},\ and\ \citenamefont {Gambetta}}]{IBM_parametric}%
  \BibitemOpen
  \bibfield  {author} {\bibinfo {author} {\bibfnamefont {D.~C.}\ \bibnamefont
  {McKay}}, \bibinfo {author} {\bibfnamefont {S.}~\bibnamefont {Filipp}},
  \bibinfo {author} {\bibfnamefont {A.}~\bibnamefont {Mezzacapo}}, \bibinfo
  {author} {\bibfnamefont {E.}~\bibnamefont {Magesan}}, \bibinfo {author}
  {\bibfnamefont {J.~M.}\ \bibnamefont {Chow}},\ and\ \bibinfo {author}
  {\bibfnamefont {J.~M.}\ \bibnamefont {Gambetta}},\ }\bibfield  {title}
  {\bibinfo {title} {\textit{Universal Gate for Fixed-Frequency Qubits via a
  Tunable Bus}},\ }\href {https://doi.org/10.1103/PhysRevApplied.6.064007}
  {\bibfield  {journal} {\bibinfo  {journal} {Phys. Rev. Appl.}\ }\textbf
  {\bibinfo {volume} {6}},\ \bibinfo {pages} {064007} (\bibinfo {year}
  {2016})}\BibitemShut {NoStop}%
\bibitem [{\citenamefont {Groszkowski}\ and\ \citenamefont
  {Koch}(2021)}]{Koch_scqubits}%
  \BibitemOpen
  \bibfield  {author} {\bibinfo {author} {\bibfnamefont {P.}~\bibnamefont
  {Groszkowski}}\ and\ \bibinfo {author} {\bibfnamefont {J.}~\bibnamefont
  {Koch}},\ }\bibfield  {title} {\bibinfo {title} {\textit{Scqubits: a Python
  Package for Superconducting Qubits}},\ }\href
  {https://doi.org/https://doi.org/10.22331/q-2021-11-17-583} {\bibfield
  {journal} {\bibinfo  {journal} {Quantum}\ }\textbf {\bibinfo {volume} {5}},\
  \bibinfo {pages} {583} (\bibinfo {year} {2021})}\BibitemShut {NoStop}%
\bibitem [{\citenamefont {Malekakhlagh}\ and\ \citenamefont
  {Magesan}(2022)}]{IBM_CR_off_res_error}%
  \BibitemOpen
  \bibfield  {author} {\bibinfo {author} {\bibfnamefont {M.}~\bibnamefont
  {Malekakhlagh}}\ and\ \bibinfo {author} {\bibfnamefont {E.}~\bibnamefont
  {Magesan}},\ }\bibfield  {title} {\bibinfo {title} {\textit{Mitigating
  Off-Resonant Error in the Cross-Resonance Gate}},\ }\href
  {https://doi.org/10.1103/PhysRevA.105.012602} {\bibfield  {journal} {\bibinfo
   {journal} {Phys. Rev. A}\ }\textbf {\bibinfo {volume} {105}},\ \bibinfo
  {pages} {012602} (\bibinfo {year} {2022})}\BibitemShut {NoStop}%
\bibitem [{Note1()}]{Note1}%
  \BibitemOpen
  \bibinfo {note} {In the dressed basis, the charge operator of the resonator
  yields other small terms, but for a constant drive amplitude, those
  additional terms are far off-resonant.}\BibitemShut {Stop}%
\bibitem [{\citenamefont {Johansson}\ \emph {et~al.}(2013)\citenamefont
  {Johansson}, \citenamefont {Nation},\ and\ \citenamefont {Nori}}]{qutip}%
  \BibitemOpen
  \bibfield  {author} {\bibinfo {author} {\bibfnamefont {J.~R.}\ \bibnamefont
  {Johansson}}, \bibinfo {author} {\bibfnamefont {P.~D.}\ \bibnamefont
  {Nation}},\ and\ \bibinfo {author} {\bibfnamefont {F.}~\bibnamefont {Nori}},\
  }\bibfield  {title} {\bibinfo {title} {\textit{QuTiP 2: A Python Framework
  for the Dynamics of Open Quantum Systems}},\ }\href
  {https://doi.org/10.1016/j.cpc.2012.11.019} {\bibfield  {journal} {\bibinfo
  {journal} {Comp. Phys. Comm.}\ }\textbf {\bibinfo {volume} {184}},\ \bibinfo
  {pages} {1234} (\bibinfo {year} {2013})}\BibitemShut {NoStop}%
\bibitem [{\citenamefont {Rigetti}\ and\ \citenamefont
  {Devoret}(2010{\natexlab{b}})}]{Devoret_CR_gate}%
  \BibitemOpen
  \bibfield  {author} {\bibinfo {author} {\bibfnamefont {C.}~\bibnamefont
  {Rigetti}}\ and\ \bibinfo {author} {\bibfnamefont {M.}~\bibnamefont
  {Devoret}},\ }\bibfield  {title} {\bibinfo {title} {\textit{Fully
  Microwave-Tunable Universal Gates in Superconducting Qubits with Linear
  Couplings and Fixed Transition Frequencies}},\ }\href
  {https://doi.org/10.1103/PhysRevB.81.134507} {\bibfield  {journal} {\bibinfo
  {journal} {Phys. Rev. B}\ }\textbf {\bibinfo {volume} {81}},\ \bibinfo
  {pages} {134507} (\bibinfo {year} {2010}{\natexlab{b}})}\BibitemShut
  {NoStop}%
\bibitem [{\citenamefont {Chow}\ \emph {et~al.}(2011)\citenamefont {Chow},
  \citenamefont {C\'orcoles}, \citenamefont {Gambetta}, \citenamefont
  {Rigetti}, \citenamefont {Johnson}, \citenamefont {Smolin}, \citenamefont
  {Rozen}, \citenamefont {Keefe}, \citenamefont {Rothwell}, \citenamefont
  {Ketchen},\ and\ \citenamefont {Steffen}}]{IBM_CR}%
  \BibitemOpen
  \bibfield  {author} {\bibinfo {author} {\bibfnamefont {J.~M.}\ \bibnamefont
  {Chow}}, \bibinfo {author} {\bibfnamefont {A.~D.}\ \bibnamefont
  {C\'orcoles}}, \bibinfo {author} {\bibfnamefont {J.~M.}\ \bibnamefont
  {Gambetta}}, \bibinfo {author} {\bibfnamefont {C.}~\bibnamefont {Rigetti}},
  \bibinfo {author} {\bibfnamefont {B.~R.}\ \bibnamefont {Johnson}}, \bibinfo
  {author} {\bibfnamefont {J.~A.}\ \bibnamefont {Smolin}}, \bibinfo {author}
  {\bibfnamefont {J.~R.}\ \bibnamefont {Rozen}}, \bibinfo {author}
  {\bibfnamefont {G.~A.}\ \bibnamefont {Keefe}}, \bibinfo {author}
  {\bibfnamefont {M.~B.}\ \bibnamefont {Rothwell}}, \bibinfo {author}
  {\bibfnamefont {M.~B.}\ \bibnamefont {Ketchen}},\ and\ \bibinfo {author}
  {\bibfnamefont {M.}~\bibnamefont {Steffen}},\ }\bibfield  {title} {\bibinfo
  {title} {\textit{Simple All-Microwave Entangling Gate for Fixed-Frequency
  Superconducting Qubits}},\ }\href
  {https://doi.org/10.1103/PhysRevLett.107.080502} {\bibfield  {journal}
  {\bibinfo  {journal} {Phys. Rev. Lett.}\ }\textbf {\bibinfo {volume} {107}},\
  \bibinfo {pages} {080502} (\bibinfo {year} {2011})}\BibitemShut {NoStop}%
\bibitem [{\citenamefont {Arute}\ \emph {et~al.}(2019)\citenamefont {Arute},
  \citenamefont {Arya}, \citenamefont {Babbush}, \citenamefont {Bacon},
  \citenamefont {Bardin}, \citenamefont {Barends}, \citenamefont {Biswas},
  \citenamefont {Boixo}, \citenamefont {Brandao}, \citenamefont {Buell},
  \citenamefont {Burkett}, \citenamefont {Chen}, \citenamefont {Chen},
  \citenamefont {Chiaro}, \citenamefont {Collins}, \citenamefont {Courtney},
  \citenamefont {Dunsworth}, \citenamefont {Farhi}, \citenamefont {Foxen},\
  and\ \citenamefont {Fowler~\textit{et al.}}}]{google_quantum_superamacy}%
  \BibitemOpen
  \bibfield  {author} {\bibinfo {author} {\bibfnamefont {F.}~\bibnamefont
  {Arute}}, \bibinfo {author} {\bibfnamefont {K.}~\bibnamefont {Arya}},
  \bibinfo {author} {\bibfnamefont {R.}~\bibnamefont {Babbush}}, \bibinfo
  {author} {\bibfnamefont {D.}~\bibnamefont {Bacon}}, \bibinfo {author}
  {\bibfnamefont {J.~C.}\ \bibnamefont {Bardin}}, \bibinfo {author}
  {\bibfnamefont {R.}~\bibnamefont {Barends}}, \bibinfo {author} {\bibfnamefont
  {R.}~\bibnamefont {Biswas}}, \bibinfo {author} {\bibfnamefont
  {S.}~\bibnamefont {Boixo}}, \bibinfo {author} {\bibfnamefont {F.~G. S.~L.}\
  \bibnamefont {Brandao}}, \bibinfo {author} {\bibfnamefont {D.~A.}\
  \bibnamefont {Buell}}, \bibinfo {author} {\bibfnamefont {B.}~\bibnamefont
  {Burkett}}, \bibinfo {author} {\bibfnamefont {Y.}~\bibnamefont {Chen}},
  \bibinfo {author} {\bibfnamefont {Z.}~\bibnamefont {Chen}}, \bibinfo {author}
  {\bibfnamefont {B.}~\bibnamefont {Chiaro}}, \bibinfo {author} {\bibfnamefont
  {R.}~\bibnamefont {Collins}}, \bibinfo {author} {\bibfnamefont
  {W.}~\bibnamefont {Courtney}}, \bibinfo {author} {\bibfnamefont
  {A.}~\bibnamefont {Dunsworth}}, \bibinfo {author} {\bibfnamefont
  {E.}~\bibnamefont {Farhi}}, \bibinfo {author} {\bibfnamefont
  {B.}~\bibnamefont {Foxen}},\ and\ \bibinfo {author} {\bibfnamefont
  {A.}~\bibnamefont {Fowler~\textit{et al.}}},\ }\bibfield  {title} {\bibinfo
  {title} {\textit{Quantum Supremacy Using a Programmable Superconducting
  Processor}},\ }\href {https://doi.org/10.1038/s41586-019-1666-5} {\bibfield
  {journal} {\bibinfo  {journal} {Nature}\ }\textbf {\bibinfo {volume} {574}},\
  \bibinfo {pages} {505} (\bibinfo {year} {2019})}\BibitemShut {NoStop}%
\bibitem [{\citenamefont {Roy}\ \emph {et~al.}(2023)\citenamefont {Roy},
  \citenamefont {Li}, \citenamefont {Kapit},\ and\ \citenamefont
  {Schuster}}]{Schuster_qutrit}%
  \BibitemOpen
  \bibfield  {author} {\bibinfo {author} {\bibfnamefont {T.}~\bibnamefont
  {Roy}}, \bibinfo {author} {\bibfnamefont {Z.}~\bibnamefont {Li}}, \bibinfo
  {author} {\bibfnamefont {E.}~\bibnamefont {Kapit}},\ and\ \bibinfo {author}
  {\bibfnamefont {D.~I.}\ \bibnamefont {Schuster}},\ }\bibfield  {title}
  {\bibinfo {title} {\textit{Two-Qutrit Quantum Algorithms on a Programmable
  Superconducting Processor}},\ }\href
  {https://doi.org/10.1103/PhysRevApplied.19.064024} {\bibfield  {journal}
  {\bibinfo  {journal} {Phys. Rev. Appl.}\ }\textbf {\bibinfo {volume} {19}},\
  \bibinfo {pages} {064024} (\bibinfo {year} {2023})}\BibitemShut {NoStop}%
\bibitem [{\citenamefont {Abad}\ \emph {et~al.}(2022)\citenamefont {Abad},
  \citenamefont {Fern\'andez-Pend\'as}, \citenamefont {Frisk~Kockum},\ and\
  \citenamefont {Johansson}}]{Goran_universal_reduction}%
  \BibitemOpen
  \bibfield  {author} {\bibinfo {author} {\bibfnamefont {T.}~\bibnamefont
  {Abad}}, \bibinfo {author} {\bibfnamefont {J.}~\bibnamefont
  {Fern\'andez-Pend\'as}}, \bibinfo {author} {\bibfnamefont {A.}~\bibnamefont
  {Frisk~Kockum}},\ and\ \bibinfo {author} {\bibfnamefont {G.}~\bibnamefont
  {Johansson}},\ }\bibfield  {title} {\bibinfo {title} {\textit{Universal
  Fidelity Reduction of Quantum Operations from Weak Dissipation}},\ }\href
  {https://doi.org/10.1103/PhysRevLett.129.150504} {\bibfield  {journal}
  {\bibinfo  {journal} {Phys. Rev. Lett.}\ }\textbf {\bibinfo {volume} {129}},\
  \bibinfo {pages} {150504} (\bibinfo {year} {2022})}\BibitemShut {NoStop}%
\bibitem [{\citenamefont {Huang}\ \emph {et~al.}(2023)\citenamefont {Huang},
  \citenamefont {Lu}, \citenamefont {Grassellino}, \citenamefont {Romanenko},
  \citenamefont {Koch},\ and\ \citenamefont {Zhu}}]{Huang_CPTP}%
  \BibitemOpen
  \bibfield  {author} {\bibinfo {author} {\bibfnamefont {Z.}~\bibnamefont
  {Huang}}, \bibinfo {author} {\bibfnamefont {Y.}~\bibnamefont {Lu}}, \bibinfo
  {author} {\bibfnamefont {A.}~\bibnamefont {Grassellino}}, \bibinfo {author}
  {\bibfnamefont {A.}~\bibnamefont {Romanenko}}, \bibinfo {author}
  {\bibfnamefont {J.}~\bibnamefont {Koch}},\ and\ \bibinfo {author}
  {\bibfnamefont {S.}~\bibnamefont {Zhu}},\ }\href@noop {} {\bibinfo {title}
  {\textit{Completely Positive Map for Noisy Driven Quantum Systems Derived by
  Keldysh Expansion}}} (\bibinfo {year} {2023}),\ \bibinfo {note}
  {\href{https://doi.org/10.48550/arXiv.2303.11491}{arXiv:2303.11491}}\BibitemShut
  {NoStop}%
\bibitem [{\citenamefont {Zhang}\ \emph {et~al.}(2019)\citenamefont {Zhang},
  \citenamefont {Lester}, \citenamefont {Gao}, \citenamefont {Jiang},
  \citenamefont {Schoelkopf},\ and\ \citenamefont
  {Girvin}}]{Yale_bilinear_coupling}%
  \BibitemOpen
  \bibfield  {author} {\bibinfo {author} {\bibfnamefont {Y.}~\bibnamefont
  {Zhang}}, \bibinfo {author} {\bibfnamefont {B.~J.}\ \bibnamefont {Lester}},
  \bibinfo {author} {\bibfnamefont {Y.~Y.}\ \bibnamefont {Gao}}, \bibinfo
  {author} {\bibfnamefont {L.}~\bibnamefont {Jiang}}, \bibinfo {author}
  {\bibfnamefont {R.~J.}\ \bibnamefont {Schoelkopf}},\ and\ \bibinfo {author}
  {\bibfnamefont {S.~M.}\ \bibnamefont {Girvin}},\ }\bibfield  {title}
  {\bibinfo {title} {\textit{Engineering Bilinear Mode Coupling in Circuit QED:
  Theory and Experiment}},\ }\href {https://doi.org/10.1103/PhysRevA.99.012314}
  {\bibfield  {journal} {\bibinfo  {journal} {Phys. Rev. A}\ }\textbf {\bibinfo
  {volume} {99}},\ \bibinfo {pages} {012314} (\bibinfo {year}
  {2019})}\BibitemShut {NoStop}%
\bibitem [{\citenamefont {Lu}\ \emph {et~al.}(2023)\citenamefont {Lu},
  \citenamefont {Maiti}, \citenamefont {Garmon}, \citenamefont {Ganjam},
  \citenamefont {Zhang}, \citenamefont {Claes}, \citenamefont {Frunzio},
  \citenamefont {Girvin},\ and\ \citenamefont
  {Schoelkopf}}]{Yao_beam_splitting}%
  \BibitemOpen
  \bibfield  {author} {\bibinfo {author} {\bibfnamefont {Y.}~\bibnamefont
  {Lu}}, \bibinfo {author} {\bibfnamefont {A.}~\bibnamefont {Maiti}}, \bibinfo
  {author} {\bibfnamefont {J.~W.~O.}\ \bibnamefont {Garmon}}, \bibinfo {author}
  {\bibfnamefont {S.}~\bibnamefont {Ganjam}}, \bibinfo {author} {\bibfnamefont
  {Y.}~\bibnamefont {Zhang}}, \bibinfo {author} {\bibfnamefont
  {J.}~\bibnamefont {Claes}}, \bibinfo {author} {\bibfnamefont
  {L.}~\bibnamefont {Frunzio}}, \bibinfo {author} {\bibfnamefont {S.~M.}\
  \bibnamefont {Girvin}},\ and\ \bibinfo {author} {\bibfnamefont {R.~J.}\
  \bibnamefont {Schoelkopf}},\ }\bibfield  {title} {\bibinfo {title}
  {\textit{High-Fidelity Parametric Beamsplitting with a Parity-Protected
  Converter}},\ }\href {https://doi.org/10.1038/s41467-023-41104-0} {\bibfield
  {journal} {\bibinfo  {journal} {Nat. Commun.}\ }\textbf {\bibinfo {volume}
  {14}},\ \bibinfo {pages} {5767} (\bibinfo {year} {2023})}\BibitemShut
  {NoStop}%
\bibitem [{\citenamefont {Motzoi}\ \emph {et~al.}(2009)\citenamefont {Motzoi},
  \citenamefont {Gambetta}, \citenamefont {Rebentrost},\ and\ \citenamefont
  {Wilhelm}}]{Wilhelm_DRAG}%
  \BibitemOpen
  \bibfield  {author} {\bibinfo {author} {\bibfnamefont {F.}~\bibnamefont
  {Motzoi}}, \bibinfo {author} {\bibfnamefont {J.~M.}\ \bibnamefont
  {Gambetta}}, \bibinfo {author} {\bibfnamefont {P.}~\bibnamefont
  {Rebentrost}},\ and\ \bibinfo {author} {\bibfnamefont {F.~K.}\ \bibnamefont
  {Wilhelm}},\ }\bibfield  {title} {\bibinfo {title} {\textit{Simple Pulses for
  Elimination of Leakage in Weakly Nonlinear Qubits}},\ }\href
  {https://doi.org/10.1103/PhysRevLett.103.110501} {\bibfield  {journal}
  {\bibinfo  {journal} {Phys. Rev. Lett.}\ }\textbf {\bibinfo {volume} {103}},\
  \bibinfo {pages} {110501} (\bibinfo {year} {2009})}\BibitemShut {NoStop}%
\bibitem [{\citenamefont {Gambetta}\ \emph {et~al.}(2011)\citenamefont
  {Gambetta}, \citenamefont {Motzoi}, \citenamefont {Merkel},\ and\
  \citenamefont {Wilhelm}}]{Wilhelm_DRAG_analy}%
  \BibitemOpen
  \bibfield  {author} {\bibinfo {author} {\bibfnamefont {J.~M.}\ \bibnamefont
  {Gambetta}}, \bibinfo {author} {\bibfnamefont {F.}~\bibnamefont {Motzoi}},
  \bibinfo {author} {\bibfnamefont {S.~T.}\ \bibnamefont {Merkel}},\ and\
  \bibinfo {author} {\bibfnamefont {F.~K.}\ \bibnamefont {Wilhelm}},\
  }\bibfield  {title} {\bibinfo {title} {\textit{Analytic Control Methods for
  High-Fidelity Unitary Operations in a Weakly Nonlinear Oscillator}},\ }\href
  {https://doi.org/10.1103/PhysRevA.83.012308} {\bibfield  {journal} {\bibinfo
  {journal} {Phys. Rev. A}\ }\textbf {\bibinfo {volume} {83}},\ \bibinfo
  {pages} {012308} (\bibinfo {year} {2011})}\BibitemShut {NoStop}%
\bibitem [{\citenamefont {Gu\'ery-Odelin}\ \emph {et~al.}(2019)\citenamefont
  {Gu\'ery-Odelin}, \citenamefont {Ruschhaupt}, \citenamefont {Kiely},
  \citenamefont {Torrontegui}, \citenamefont {Mart\'{\i}nez-Garaot},\ and\
  \citenamefont {Muga}}]{Muga_adiabaticity_shortcut}%
  \BibitemOpen
  \bibfield  {author} {\bibinfo {author} {\bibfnamefont {D.}~\bibnamefont
  {Gu\'ery-Odelin}}, \bibinfo {author} {\bibfnamefont {A.}~\bibnamefont
  {Ruschhaupt}}, \bibinfo {author} {\bibfnamefont {A.}~\bibnamefont {Kiely}},
  \bibinfo {author} {\bibfnamefont {E.}~\bibnamefont {Torrontegui}}, \bibinfo
  {author} {\bibfnamefont {S.}~\bibnamefont {Mart\'{\i}nez-Garaot}},\ and\
  \bibinfo {author} {\bibfnamefont {J.~G.}\ \bibnamefont {Muga}},\ }\bibfield
  {title} {\bibinfo {title} {\textit{Shortcuts to Adiabaticity: Concepts,
  Methods, and Applications}},\ }\href
  {https://doi.org/10.1103/RevModPhys.91.045001} {\bibfield  {journal}
  {\bibinfo  {journal} {Rev. Mod. Phys.}\ }\textbf {\bibinfo {volume} {91}},\
  \bibinfo {pages} {045001} (\bibinfo {year} {2019})}\BibitemShut {NoStop}%
\bibitem [{\citenamefont {Sank}\ \emph {et~al.}(2016)\citenamefont {Sank},
  \citenamefont {Chen}, \citenamefont {Khezri}, \citenamefont {Kelly},
  \citenamefont {Barends}, \citenamefont {Campbell}, \citenamefont {Chen},
  \citenamefont {Chiaro}, \citenamefont {Dunsworth}, \citenamefont {Fowler},
  \citenamefont {Jeffrey}, \citenamefont {Lucero}, \citenamefont {Megrant},
  \citenamefont {Mutus}, \citenamefont {Neeley}, \citenamefont {Neill},
  \citenamefont {O'Malley}, \citenamefont {Quintana}, \citenamefont {Roushan},
  \citenamefont {Vainsencher}, \citenamefont {White}, \citenamefont {Wenner},
  \citenamefont {Korotkov},\ and\ \citenamefont {Martinis}}]{Martinis_leakage}%
  \BibitemOpen
  \bibfield  {author} {\bibinfo {author} {\bibfnamefont {D.}~\bibnamefont
  {Sank}}, \bibinfo {author} {\bibfnamefont {Z.}~\bibnamefont {Chen}}, \bibinfo
  {author} {\bibfnamefont {M.}~\bibnamefont {Khezri}}, \bibinfo {author}
  {\bibfnamefont {J.}~\bibnamefont {Kelly}}, \bibinfo {author} {\bibfnamefont
  {R.}~\bibnamefont {Barends}}, \bibinfo {author} {\bibfnamefont
  {B.}~\bibnamefont {Campbell}}, \bibinfo {author} {\bibfnamefont
  {Y.}~\bibnamefont {Chen}}, \bibinfo {author} {\bibfnamefont {B.}~\bibnamefont
  {Chiaro}}, \bibinfo {author} {\bibfnamefont {A.}~\bibnamefont {Dunsworth}},
  \bibinfo {author} {\bibfnamefont {A.}~\bibnamefont {Fowler}}, \bibinfo
  {author} {\bibfnamefont {E.}~\bibnamefont {Jeffrey}}, \bibinfo {author}
  {\bibfnamefont {E.}~\bibnamefont {Lucero}}, \bibinfo {author} {\bibfnamefont
  {A.}~\bibnamefont {Megrant}}, \bibinfo {author} {\bibfnamefont
  {J.}~\bibnamefont {Mutus}}, \bibinfo {author} {\bibfnamefont
  {M.}~\bibnamefont {Neeley}}, \bibinfo {author} {\bibfnamefont
  {C.}~\bibnamefont {Neill}}, \bibinfo {author} {\bibfnamefont {P.~J.~J.}\
  \bibnamefont {O'Malley}}, \bibinfo {author} {\bibfnamefont {C.}~\bibnamefont
  {Quintana}}, \bibinfo {author} {\bibfnamefont {P.}~\bibnamefont {Roushan}},
  \bibinfo {author} {\bibfnamefont {A.}~\bibnamefont {Vainsencher}}, \bibinfo
  {author} {\bibfnamefont {T.}~\bibnamefont {White}}, \bibinfo {author}
  {\bibfnamefont {J.}~\bibnamefont {Wenner}}, \bibinfo {author} {\bibfnamefont
  {A.~N.}\ \bibnamefont {Korotkov}},\ and\ \bibinfo {author} {\bibfnamefont
  {J.~M.}\ \bibnamefont {Martinis}},\ }\bibfield  {title} {\bibinfo {title}
  {\textit{Measurement-Induced State Transitions in a Superconducting Qubit:
  Beyond the Rotating Wave Approximation}},\ }\href
  {https://doi.org/10.1103/PhysRevLett.117.190503} {\bibfield  {journal}
  {\bibinfo  {journal} {Phys. Rev. Lett.}\ }\textbf {\bibinfo {volume} {117}},\
  \bibinfo {pages} {190503} (\bibinfo {year} {2016})}\BibitemShut {NoStop}%
\bibitem [{\citenamefont {Gambetta}\ \emph {et~al.}(2006)\citenamefont
  {Gambetta}, \citenamefont {Blais}, \citenamefont {Schuster}, \citenamefont
  {Wallraff}, \citenamefont {Frunzio}, \citenamefont {Majer}, \citenamefont
  {Devoret}, \citenamefont {Girvin},\ and\ \citenamefont
  {Schoelkopf}}]{Schoelkopf_driven_cavity_dephasing}%
  \BibitemOpen
  \bibfield  {author} {\bibinfo {author} {\bibfnamefont {J.}~\bibnamefont
  {Gambetta}}, \bibinfo {author} {\bibfnamefont {A.}~\bibnamefont {Blais}},
  \bibinfo {author} {\bibfnamefont {D.~I.}\ \bibnamefont {Schuster}}, \bibinfo
  {author} {\bibfnamefont {A.}~\bibnamefont {Wallraff}}, \bibinfo {author}
  {\bibfnamefont {L.}~\bibnamefont {Frunzio}}, \bibinfo {author} {\bibfnamefont
  {J.}~\bibnamefont {Majer}}, \bibinfo {author} {\bibfnamefont {M.~H.}\
  \bibnamefont {Devoret}}, \bibinfo {author} {\bibfnamefont {S.~M.}\
  \bibnamefont {Girvin}},\ and\ \bibinfo {author} {\bibfnamefont {R.~J.}\
  \bibnamefont {Schoelkopf}},\ }\bibfield  {title} {\bibinfo {title}
  {\textit{Qubit-Photon Interactions in a Cavity: Measurement-Induced Dephasing
  and Number Splitting}},\ }\href {https://doi.org/10.1103/PhysRevA.74.042318}
  {\bibfield  {journal} {\bibinfo  {journal} {Phys. Rev. A}\ }\textbf {\bibinfo
  {volume} {74}},\ \bibinfo {pages} {042318} (\bibinfo {year}
  {2006})}\BibitemShut {NoStop}%
\bibitem [{\citenamefont {Gyenis}\ \emph {et~al.}(2021)\citenamefont {Gyenis},
  \citenamefont {Di~Paolo}, \citenamefont {Koch}, \citenamefont {Blais},
  \citenamefont {Houck},\ and\ \citenamefont
  {Schuster}}]{Gyenis_protected_review}%
  \BibitemOpen
  \bibfield  {author} {\bibinfo {author} {\bibfnamefont {A.}~\bibnamefont
  {Gyenis}}, \bibinfo {author} {\bibfnamefont {A.}~\bibnamefont {Di~Paolo}},
  \bibinfo {author} {\bibfnamefont {J.}~\bibnamefont {Koch}}, \bibinfo {author}
  {\bibfnamefont {A.}~\bibnamefont {Blais}}, \bibinfo {author} {\bibfnamefont
  {A.~A.}\ \bibnamefont {Houck}},\ and\ \bibinfo {author} {\bibfnamefont
  {D.~I.}\ \bibnamefont {Schuster}},\ }\bibfield  {title} {\bibinfo {title}
  {\textit{Moving beyond the Transmon: Noise-Protected Superconducting Quantum
  Circuits }},\ }\href {https://doi.org/10.1103/PRXQuantum.2.030101} {\bibfield
   {journal} {\bibinfo  {journal} {PRX Quantum}\ }\textbf {\bibinfo {volume}
  {2}},\ \bibinfo {pages} {030101} (\bibinfo {year} {2021})}\BibitemShut
  {NoStop}%
\bibitem [{Note2()}]{Note2}%
  \BibitemOpen
  \bibinfo {note} {More accurately, this coefficient is derived as $\chi
  _{L(R)} = 2\protect \tilde {g}^2_{L(R)}\eta _{L(R)}/[(\protect \bar {\omega
  }_{L(R)}-\protect \bar {\omega }_C)(\protect \bar {\omega }_{L(R)}-\protect
  \bar {\omega }_C+\eta _{L(R)})]$ \cite {Blais_cqed_review}.}\BibitemShut
  {Stop}%
\bibitem [{\citenamefont {Kafri}\ \emph {et~al.}(2017)\citenamefont {Kafri},
  \citenamefont {Quintana}, \citenamefont {Chen}, \citenamefont {Shabani},
  \citenamefont {Martinis},\ and\ \citenamefont {Neven}}]{Dvir_BO}%
  \BibitemOpen
  \bibfield  {author} {\bibinfo {author} {\bibfnamefont {D.}~\bibnamefont
  {Kafri}}, \bibinfo {author} {\bibfnamefont {C.}~\bibnamefont {Quintana}},
  \bibinfo {author} {\bibfnamefont {Y.}~\bibnamefont {Chen}}, \bibinfo {author}
  {\bibfnamefont {A.}~\bibnamefont {Shabani}}, \bibinfo {author} {\bibfnamefont
  {J.~M.}\ \bibnamefont {Martinis}},\ and\ \bibinfo {author} {\bibfnamefont
  {H.}~\bibnamefont {Neven}},\ }\bibfield  {title} {\bibinfo {title}
  {\textit{Tunable Inductive Coupling of Superconducting Qubits in the Strongly
  Nonlinear Regime}},\ }\href {https://doi.org/10.1103/PhysRevA.95.052333}
  {\bibfield  {journal} {\bibinfo  {journal} {Phys. Rev. A}\ }\textbf {\bibinfo
  {volume} {95}},\ \bibinfo {pages} {052333} (\bibinfo {year}
  {2017})}\BibitemShut {NoStop}%
\bibitem [{\citenamefont {Huang}(2021)}]{Huang_dissertation}%
  \BibitemOpen
  \bibfield  {author} {\bibinfo {author} {\bibfnamefont {Z.}~\bibnamefont
  {Huang}},\ }\emph {\bibinfo {title} {\textit{Noise Engineering and Mitigation
  in Superconducting Circuits}}},\ \href@noop {} {Ph.D. thesis},\ \bibinfo
  {school} {Northwestern University} (\bibinfo {year} {2021})\BibitemShut
  {NoStop}%
\end{thebibliography}
\end{document}